\documentstyle [a4,11pt,axodraw]{article}

\begin{document}
\begin{titlepage}
\hfill FTUAM 96-7 \\ [-5mm]

\hfill 21-nov-1996 \\

\begin{center}
{\huge Pomerons and Jet Events at HERA} \\ [12mm]
\Large J. Vermaseren \\ [3mm]
NIKHEF \\
P.O.Box 41882 \\
NL-1009DB, Amsterdam \\ [5mm]
\Large F. Barreiro, L. Labarga and F. J. Yndur\'ain \\ [3mm]
Departamento de F\'\i sica Te\'orica, C-XI, \\
Universidad Aut\'onoma de Madrid, \\
Canto Blanco,\\
E-28034 Madrid, SPAIN. \\ [10mm]
\end{center}
 
\begin{abstract}\noindent
We study two and three jet events with a large rapidity gap at HERA.
Unlike in the Ingelman-Schlein approach we do not adscribe a 
structure to the Pomeron. Instead, 
the coupling of the Pomeron to quarks or gluons is 
taken pointlike, which makes the model easy to test:
 the only degrees of freedom are the coupling constants of the Pomeron to 
the quarks or the gluons and a cutoff procedure to keep the Pomeron-gluon 
coupling well behaved.
\end{abstract}
\end{titlepage}

\section{Introduction}

The prediction of the Pomeron as a pseudo particle has inspired a large 
number of experiments in the 60's and 70's. Although effective parameters 
could be determined under the assumption that the particle exists, its 
existence was never shown unambiguously. Recently the so-called `large 
rapidity gap' events at HERA \cite{zrap} have caused new
 excitement about the Pomeron 
hypothesis. These events show jet activity far away from the beam pipe, but 
there is no trace of a color string between these jets and the remnants of 
the proton. Actually it is not excluded that in many or 
even most cases the proton remains intact. Hence the events carry the 
signature of a diffractive mechanism in which a color neutral object reacts 
with the virtual photon that comes from the electron. The outcoming events 
can be classified by the number of jets they show. This is similar to 
$e^-e^+\rightarrow hadrons$ with which these events can be shown to have 
much in common. We will comment on this in the results section.

Thus far most theoretical analysis has concerned the two jet events. 
Various models have been tried, ranging from a full calculation of a gluon 
pair exchange~\cite{DandL1} and \cite{NIKZAK}, to a
 phenomenological description of Pomeron structure 
functions~\cite{Pompyt}~\cite{Jung} and even to a gluon photon fusion 
scheme in which the resulting quark pair somehow sheds its color and leaves 
it behind inside the proton fragments~\cite{Buchmueller}. In this paper we 
will produce the Pomeron with a classically parameterized diffractive 
mechanism~\cite{DandL2}. Having produced it, we will couple it to quarks 
and/or gluons with a pointlike coupling. This has been tried before for two 
jet events, but then in the case of a small $Q^2$ of the 
photon~\cite{point}; see also ref~\cite{bartels}. In this paper we are 
however interested in the region of large $Q^2$ in which the outgoing 
electron gets observed clearly inside the detector. In addition we will 
have a look at three jet events. The aim of the paper is then to see how 
well the events can be used to determine the coupling of the Pomeron to the 
quarks (and/or gluons).

There are various motivations for this model. First it is the simplest 
model one can think of, and hence it is necessary to have at least a look 
at it. Another motivation has to do with the fact that the model contains
explicit calculations for three-jet production allowing us to make simple
predictions for broadening effects in the multihadronic diffractive final 
states. In addition the matrix elements can be implemented in a Monte Carlo
programme using the Lund fragmentation scheme. Thus, our model can be directly
compared to experimental data. In fact, as it will be shown in a forthcoming
publication \cite{expzeus}, the model gives a good description of the data.



\section{The computation}

Considering the fact that we insist on seeing the electron, we can treat it 
in the normal way. This means that it is just another particle in the 
trace calculations, and we do not have to take extra precautions with 
respect to numerical stability of either the matrix element or the 
kinematics. Because of this there is absolutely no need to use an 
equivalent photon approximation. If the electron were to stay in the 
beampipe the situation would be different. In the forward direction terms 
of order $m_e^2/s$ are important and seemingly leading terms cancel each 
other to a high degree of accuracy. This not being the case we can work out 
the trace of the Feynman diagrams by brute force using the program 
FORM~\cite{form} and evaluate the resulting FORTRAN code without much 
worry. The two jet diagrams of figure 1 are easy to compute.
\begin{center}
\begin{picture}(240,80)(0,0)
\SetScale{0.8}
\ArrowLine(60,90)(100,90)
\ArrowLine(100,90)(140,90)
\Photon(100,90)(100,65){1.5}{3}
\ArrowLine(100,65)(140,65)
\ArrowLine(100,25)(100,65)
\ArrowLine(140,25)(100,25)
\ZigZag(60,25)(100,25){3}{5}
\Vertex(100,90){1}
\Vertex(100,65){1}
\Vertex(100,25){1}

\ArrowLine(160,90)(200,90)
\ArrowLine(200,90)(240,90)
\Photon(200,90)(200,65){1.5}{3}
\ArrowLine(240,65)(200,65)
\ArrowLine(200,65)(200,25)
\ArrowLine(200,25)(240,25)
\ZigZag(160,25)(200,25){3}{5}
\Vertex(200,90){1}
\Vertex(200,65){1}
\Vertex(200,25){1}
\end{picture} \\
\parbox[]{350pt}{{\sl Figure 1}:
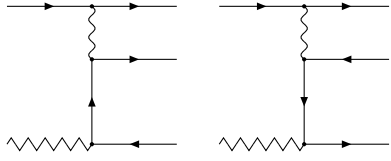
	The 2 diagrams for Pomeron induced two jet production. The Pomeron is 
    denoted by the zigzag line at the left bottom of each diagram}
\end{center}
In this case there is only one `new' parameter. The 
strength of the coupling of the Pomeron to the quarks: $g_{Pqq}$. Basically 
the normalization of the reaction fixes this constant. The cross section 
is easily calculated. If one were to ignore the polarization of the photon 
and the 'mass' of the Pomeron the matrix element of the subprocess
\[\gamma (q)+{\rm P(p)}\rightarrow \bar{q}(k_1)+q(k_2)\]
(in brackets the momenta of the particles) then its cross section, denoted by 
$\hat{\sigma}_{\gamma{\rm P}}$
would be
\[d\hat{\sigma}_{\gamma{\rm P}}=
\frac{2g^2_{{}_{Pqq}}\alpha}{\pi}\,\frac{s-q^2}{ut}\delta (p+q-k_1-k_2)
\frac{d^3k_1}{2k_{10}}\,\frac{d^3k_2}{2k_{20}}.\]
Here $t,u$ are the standard Mandelstam variables ($t$ is defined by P and 
$k_1$), and $g_{{}_{Pqq}}$ is defined by the effective interaction Lagrangian ${\cal 
L}_{\rm eff} = g_{{}_{Pqq}}\bar{q}qP$. In practice we did not ignore the 
polarization of the photon and the 'mass' of the Pomeron. Hence the formula 
for the full process is a little longer:
\begin{eqnarray}
	d\sigma_{eP} & = & \frac{d^3k_1}{2E_1}\frac{d^3k_2}{2E_2}
			\frac{d^3p_e^{out}}{2E_e^{out}}\delta^4(p_e+P-k_1-k_2-p_e^{out})
			(-q^2\frac{(M_X^2-q^2)^2+P^2}{ut}
			+P^2 (q^2)^2 (\frac{1}{u^2}+\frac{1}{t^2})
	\nonumber \\ &&
			+8 P\cdot p_e(\frac{1}{u}+\frac{1}{t})(
			\frac{p_e\cdot k_1\ (P^2-u)}{t}
			 + \frac{p_e\cdot k_2\ (P^2-t)}{u})
	\nonumber \\ &&
			-8 (P\cdot p_e)^2 \frac{M_X^2}{ut}
			-8 P^2\ p_e\cdot k_1\ p_e\cdot k_2
				\ (\frac{1}{u}+\frac{1}{t})^2 )
			\frac{g^2_{{}_{Pqq}}\alpha^2}{(q^2)^2\pi^3(s-P^2)}
\end{eqnarray}
We use here $M_X$ for the diquark invariant mass, $p_e$ for the incoming 
electron and $p_e^{out}$ for the outgoing electron.

The three parton diagrams, responsible mostly for the three jet events, are 
given in figure 2.
Here the formulae are much more complicated, but with the use of FORM we 
obtain an expression of about 800 terms which can be written into a 
file in such a way that it can be included into a FORTRAN program directly.
We have now another `new' parameter: $g_{{}_{Pgg}}$, the coupling of the Pomeron 
to two gluons. Its size can be fixed by looking at the ratio between two 
and three jet events. It should be noted that because the Pomeron is taken 
to be a scalar particle and because we assumed that the quarks are 
massless, the six quark-coupling induced diagrams do not interfere with the 
two gluon-coupling induced diagrams.
\begin{center}
\begin{picture}(320,240)(0,0)
\SetScale{0.8}
\ArrowLine(60,290)(100,290)
\ArrowLine(100,290)(140,290)
\Photon(100,290)(100,265){1.5}{3}
\Vertex(100,290){1}
\ArrowLine(115,265)(140,265)
\ArrowLine(100,265)(115,265)
\Gluon(115,265)(140,245){-3}{4}
\Vertex(115,265){1}
\ArrowLine(100,225)(100,265)
\Vertex(100,265){1}
\ArrowLine(140,225)(100,225)
\ZigZag(60,225)(100,225){3}{5}
\Vertex(100,225){1}

\ArrowLine(160,290)(200,290)
\ArrowLine(200,290)(240,290)
\Photon(200,290)(200,265){1.5}{3}
\Vertex(200,290){1}
\ArrowLine(200,265)(240,265)
\ArrowLine(200,245)(200,265)
\Gluon(200,245)(240,245){3}{5}
\ArrowLine(200,225)(200,245)
\Vertex(200,245){1}
\Vertex(200,265){1}
\ArrowLine(240,225)(200,225)
\ZigZag(160,225)(200,225){3}{5}
\Vertex(200,225){1}

\ArrowLine(260,290)(300,290)
\ArrowLine(300,290)(340,290)
\Photon(300,290)(300,265){1.5}{3}
\Vertex(300,290){1}
\ArrowLine(300,265)(340,265)
\ArrowLine(300,225)(300,265)
\Vertex(300,265){1}
\ArrowLine(340,225)(315,225)
\ArrowLine(315,225)(300,225)
\Gluon(315,225)(340,245){3}{4}
\Vertex(315,225){1}
\ZigZag(260,225)(300,225){3}{5}
\Vertex(300,225){1}

\ArrowLine(60,190)(100,190)
\ArrowLine(100,190)(140,190)
\Photon(100,190)(100,165){1.5}{3}
\Vertex(100,190){1}
\ArrowLine(140,165)(100,165)
\ArrowLine(100,165)(100,125)
\Vertex(100,165){1}
\ArrowLine(115,125)(140,125)
\ArrowLine(100,125)(115,125)
\Gluon(115,125)(140,145){3}{4}
\Vertex(115,125){1}
\ZigZag(60,125)(100,125){3}{5}
\Vertex(100,125){1}

\ArrowLine(160,190)(200,190)
\ArrowLine(200,190)(240,190)
\Photon(200,190)(200,165){1.5}{3}
\Vertex(200,190){1}
\ArrowLine(240,165)(200,165)
\ArrowLine(200,165)(200,145)
\ArrowLine(200,145)(200,125)
\ArrowLine(200,125)(240,125)
\Gluon(200,145)(240,145){3}{5}
\Vertex(200,145){1}
\Vertex(200,165){1}
\ZigZag(160,125)(200,125){3}{5}
\Vertex(200,125){1}

\ArrowLine(260,190)(300,190)
\ArrowLine(300,190)(340,190)
\Photon(300,190)(300,165){1.5}{3}
\Vertex(300,190){1}
\ArrowLine(340,165)(315,165)
\ArrowLine(315,165)(300,165)
\ArrowLine(300,165)(300,125)
\ArrowLine(300,125)(340,125)
\Gluon(315,165)(340,145){-3}{4}
\Vertex(315,165){1}
\Vertex(300,165){1}
\ZigZag(260,125)(300,125){3}{5}
\Vertex(300,125){1}

\ArrowLine(110,90)(150,90)
\ArrowLine(150,90)(190,90)
\Photon(150,90)(150,65){1.5}{3}
\Vertex(150,90){1}
\ArrowLine(150,65)(180,65)
\ArrowLine(150,45)(150,65)
\ArrowLine(180,45)(150,45)
\Vertex(150,65){1}
\Gluon(150,25)(150,45){3}{3}
\Gluon(150,25)(180,25){-3}{4}
\Vertex(150,45){1}
\ZigZag(110,25)(150,25){3}{5}
\Vertex(150,25){1}

\ArrowLine(210,90)(250,90)
\ArrowLine(250,90)(290,90)
\Photon(250,90)(250,65){1.5}{3}
\Vertex(250,90){1}
\ArrowLine(280,65)(250,65)
\ArrowLine(250,65)(250,45)
\ArrowLine(250,45)(280,45)
\Vertex(250,65){1}
\Gluon(250,25)(250,45){3}{3}
\Gluon(250,25)(280,25){-3}{4}
\Vertex(250,45){1}
\ZigZag(210,25)(250,25){3}{5}
\Vertex(250,25){1}
\end{picture} \\
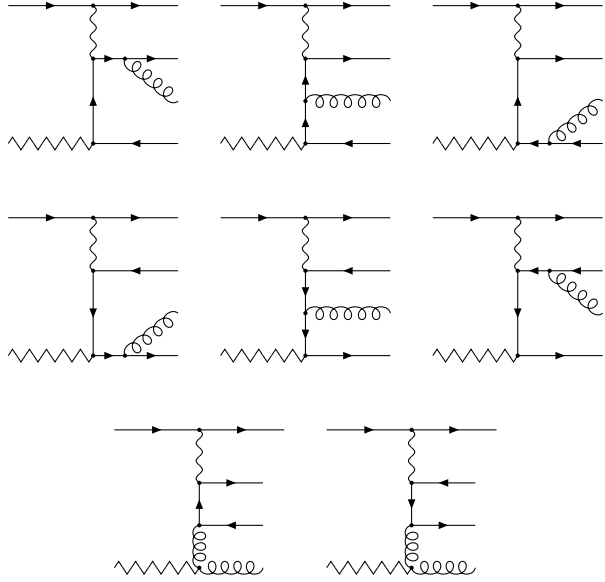
{\sl Figure 2}: The 6+2 diagrams for Pomeron induced three jet production
\end{center}

When the Pomeron is coupled to two gluons we use a $g_{Pgg}F^{\mu\nu} 
F_{\mu\nu}$ coupling. This presents some problems. Because this coupling 
causes an unphysical high energy behaviour, due to the extra dimension, 
this is `repaired' by a form factor. For this form factor we took the 
expression 
$1/(M^2-t_3)^{1/2}$ in which $t_3$ is the invariant mass of the internal gluon. 
It has a negative value. $M^2$ is a free parameter which is expected to be 
of the order of $1 GeV^2$. Hence we used this value. The presence of the 
formfactor is noticeable in the differential cross section with respect 
to $M_X$ which is the invariant mass of the three jet system. In the case 
that we do not use the form factor we get a clearly unphysical behaviour 
for large values of $M_X$. It is roughly constant until phase space cuts it 
off. With the formfactor a more physical behaviour is obtained and the mass 
distribution looks similar to the corresponding one for the 6 diagrams with 
the quark coupling.

For the production of the Pomeron we used a phenomenological fit that has 
been obtained by Donnachie and Landshoff~\cite{DandL2}. This way the 
Pomeron flux in a proton can be written as:
\begin{eqnarray}
	f_{P}(x_P,t) & = & \frac{9\beta_0^2}{4\pi^2}(F_1(t))^2
			x_P^{-2\alpha(t)+1}
\end{eqnarray}
with
\begin{eqnarray}
	F_1(t) & = & \frac{4m_p^2-At}{4m_p^2-t}(1-\frac{t}{B})^{-2}
	\nonumber \\
	A & = & 2.8
	\nonumber \\
	B & = & 0.7
	\nonumber \\
	\beta_0 & = & 3.24 GeV^{-2}
	\nonumber \\
	\alpha(t) & = & 1+\epsilon+\alpha't
	\nonumber \\
	\epsilon & = & 0.085
	\nonumber \\
	\alpha' & = & 0.25
\end{eqnarray}
It is of course rather easy to implement such a flux factor, provided that 
the integrations over phase space are done numerically. It is equally possible
to change the intercept $\epsilon$ to a value close to $0.2$ as suggested by
recent HERA data.

Next we need to do the integration. This is done by means of a Monte Carlo 
program VEGAS~\cite{vegas}. It is a so-called self learning program that 
integrates over a unit cube of a given number of dimensions. Hence we have 
to transform points in such a cube into points in phase space. The routines 
that do this are called the kinematics routines. In the current calculation 
we use some `standard' two and three body decay routines. Both routines 
assume that the `decay' takes place in the CM frame of the decaying object. 
This means that at various points we need some trivial Lorenz 
transformations. The two body decay routine is used for the reaction $e^- P 
\rightarrow e^- X$ in which $P$ is the Pomeron and $X$ represents the 2 or 
3 jet system. It is also used for the subreaction $\gamma^* P \rightarrow 
2$ jets. The three body decay routine first selects two Dalitz variables 
and then supplements them with the proper CM angles. It is used for the 
subreaction $\gamma^* P \rightarrow 3$ jets. Because everything happens at 
large angles such routines are sufficient.

Of course integration would be meaningless if we would not apply acceptance 
cuts. One has to worry about two sets of cuts. In the first set we try, at 
the parton level, to imitate the experimental analysis. In this case we do 
as if the partons are the jets of the experiment and we copy the analysis 
cuts of the experiment. In the second case we prepare the program for event 
generation allowing the experimental analysis programs to apply 
fragmentation into hadrons. We can then do the analysis at this level. 
Alternatively the resulting particles can be sent to the detector emulator 
programs. This last step would be part of a full experimental analysis. In 
the case of hadronization the cuts have to be stricter, because the 
fragmentation will make things less clear. There will be events that pass 
the analysis cuts this way that did not pass them in the theoretical 
analysis, et vice versa. In general the last method is superior, but it 
requires the application of the full experimental detector simulators. It 
also requires many more computer resources. The first method however is 
usually sufficient to give a good idea of what is happening, how 
distributions will differ between various models, etc. The study of the 
hadronization effects, without using the detector emulators, gives already 
a good indication of whether things will change much in a full 
analysis.
The cuts we use in the calculations at the parton level are:

\begin{itemize}
\item Rapitity($\eta$) of the scattered electron: $> -3.7$.
\item Energy of the scattered electron: $> 5$GeV.
\item $Q^2$ of the electron: $> 3$GeV${}^2$.
\item Mass of the outcoming hadron system($M_X$): $> 3$GeV.
\item Angle of the hadron system: $\cos\theta < 0.9$.
\item Rapidity of the individual partons: $< 3$.
\item Transverse energy of the individual partons: $> 0.2$GeV.
\item $M^2_{jj}/M^2_X \ge Y_{min} = 0.05$.
\item Beam pipe cut: Each parton has $\theta > 0.1$rad.
\end{itemize}

For the calculation that involves the hadronization we used

\begin{itemize}
\item For the jets apply the cuts for the individual partons.
\item Only particles with $\theta > 0.1$rad are used.
\item There should be at least 4 particles.
\item All particles should have a rapidity $< 2.5$.
\item Angle of the hadron system: $\cos\theta < 0.75$.
\item $Q^2$ of the electron: $> 10$GeV${}^2$.
\item Mass of the outcoming hadron system($M_X$): $> 5$GeV.
\end{itemize}

The hadronization was performed with the program JETSET 7.4~\cite{jetset}.
The parton flavors were selected to be d:u:s = 1:4:1 and the decays of some 
of the hadrons (mainly the strange ones) were suppressed.

Because we use a Monte Carlo integration method, it is easy to make a large 
variety of histograms. In addition we can use an additional set of programs 
after the integration that makes the program behave like an event generator 
in which the events all have weight one. Hence the programs are 
suitable for a detailed experimental analysis. The number for the total 
cross section can usually be obtained with an accuracy of a few tenth of a 
percent during the integration phase.

\section{Results}

In diffractive $e-p$ events, the absence of the proton remnant considerably
simplifies the analysis of the multihadronic final state. If interpreted in    
terms of $\gamma^*-P$ collisions, the Pomeron fractional momentum $\xi$,
with respect to the proton beam, can be easily obtained from the relation
\begin{equation}
\xi=\frac{M_X^2+Q^2}{M_X^2+W^2}
\end{equation}
where $M_X^2$, $Q^2$ and $W^2$ stand for the usual DIS kinematic variables.\\
We propose to establish an analogy between $\gamma^*-P$ scattering at a centre
of mass given by $M_X$ and $e^+e^-$ annihilation at a centre of mass energy
 $\sqrt{s}$. In this section, we study the dependence of shape variables
describing the multihadronic final states as a function of $M_X$, in order to
asses the importance of the two different broadening mechanisms in the model
\begin{itemize}
\item[$\clubsuit$] gluon radiation
\item[$\clubsuit$] Pomeron splitting into two gluons
\end{itemize}

To $O(\alpha_s)$, the cross-sections for the three types of processes considered
in this paper, within the cuts described in the previous section
and for a fixed value of the strong coupling constant $\alpha_s=0.25$, are:
\begin{itemize}
\item  $(43.83\pm 0.08)\cdot \ g^2_{{}_{Pqq}}~nb$ for two-jet production due
to the direct coupling of the Pomeron to quark pairs
\item  $(3.240\pm 0.003)\cdot \ g^2_{{}_{Pqq}}~nb$ for three-jet production 
from gluon bremstrahlung to the procees above
\item  $(0.2705\pm 0.0009)\cdot \ g^2_{{}_{Pgg}}~nb$ for three-jet production
from the Pomeron coupling to gluon pairs  
\end{itemize}

In Fig. 3a we show $\frac{1}{\sigma}\frac{d\ \sigma}{d\ M_X}$ for these three
reactions. Notice that three-jet processes through the Pomeron coupling to
gluons tend to give rise to larger masses then those coming from bremstrahlung.
In Fig. 3b we show the thrust \cite{thrust} distribution for three parton
final states in the $\gamma-P$ centre of mass frame. Again, 
 the tail in the thrust distribution associated with
final states due to Pomeron coupling to two gluons, dashed line,
 is harder than that associated to
bremstrahlung from the Pomeron to two quarks coupling, solid line. In fact, the
latter is in rather good agreement with the analytical expression derived by
Ellis, Gaillard and Ross for $e^+e^-$ annihilation \cite{ellis}.

\begin{center}\begin{picture}(440,400)(0,0)
%
%
\SetOffset(40,225)
\SetWidth{1.0}\BBox(0,0)(160,160)\SetWidth{0.5}
\LinAxis(0,0)(160,0)(8,5,2.5,1,1)
\LinAxis(0,160)(160,160)(8,5,-2.5,1,1)
\LogAxis(0,0)(0,160)(5,-2.5,3,1)
\LogAxis(160,0)(160,160)(5,2.5,3,1)
\Text(17,-6)[]{10}
\Text(37,-6)[]{20}
\Text(57,-6)[]{30}
\Text(77,-6)[]{40}
\Text(97,-6)[]{50}
\Text(117,-6)[]{60}
\Text(137,-6)[]{70}
\Text(157,-6)[]{80}
\Text(80,-18)[]{$M_X$}
\Text(-12,144)[]{$10^{-1}$}
\Text(-12,112)[]{$10^{-2}$}
\Text(-12,80)[]{$10^{-3}$}
\Text(-12,48)[]{$10^{-4}$}
\Text(-12,16)[]{$10^{-5}$}
\rText(-33,80)[][l]{$\frac{1}{\sigma}\frac{d\ \sigma}{d\ M_X}$}
\SetWidth{0.3}
\DashLine(0.000000,156.847649)(4.000000,156.847649){0.6}
\DashLine(4.000000,156.847649)(4.000000,146.778544){0.6}
\DashLine(4.000000,146.778544)(8.000000,146.778544){0.6}
\DashLine(8.000000,146.778544)(8.000000,137.330817){0.6}
\DashLine(8.000000,137.330817)(12.000000,137.330817){0.6}
\DashLine(12.000000,137.330817)(12.000000,129.172296){0.6}
\DashLine(12.000000,129.172296)(16.000000,129.172296){0.6}
\DashLine(16.000000,129.172296)(16.000000,121.509333){0.6}
\DashLine(16.000000,121.509333)(20.000000,121.509333){0.6}
\DashLine(20.000000,121.509333)(20.000000,114.748742){0.6}
\DashLine(20.000000,114.748742)(24.000000,114.748742){0.6}
\DashLine(24.000000,114.748742)(24.000000,108.023989){0.6}
\DashLine(24.000000,108.023989)(28.000000,108.023989){0.6}
\DashLine(28.000000,108.023989)(28.000000,102.014662){0.6}
\DashLine(28.000000,102.014662)(32.000000,102.014662){0.6}
\DashLine(32.000000,102.014662)(32.000000,96.956480){0.6}
\DashLine(32.000000,96.956480)(36.000000,96.956480){0.6}
\DashLine(36.000000,96.956480)(36.000000,91.656454){0.6}
\DashLine(36.000000,91.656454)(40.000000,91.656454){0.6}
\DashLine(40.000000,91.656454)(40.000000,86.083438){0.6}
\DashLine(40.000000,86.083438)(44.000000,86.083438){0.6}
\DashLine(44.000000,86.083438)(44.000000,81.518077){0.6}
\DashLine(44.000000,81.518077)(48.000000,81.518077){0.6}
\DashLine(48.000000,81.518077)(48.000000,79.289728){0.6}
\DashLine(48.000000,79.289728)(52.000000,79.289728){0.6}
\DashLine(52.000000,79.289728)(52.000000,73.654808){0.6}
\DashLine(52.000000,73.654808)(56.000000,73.654808){0.6}
\DashLine(56.000000,73.654808)(56.000000,68.392495){0.6}
\DashLine(56.000000,68.392495)(60.000000,68.392495){0.6}
\DashLine(60.000000,68.392495)(60.000000,68.826825){0.6}
\DashLine(60.000000,68.826825)(64.000000,68.826825){0.6}
\DashLine(64.000000,68.826825)(64.000000,64.320729){0.6}
\DashLine(64.000000,64.320729)(68.000000,64.320729){0.6}
\DashLine(68.000000,64.320729)(68.000000,59.820923){0.6}
\DashLine(68.000000,59.820923)(72.000000,59.820923){0.6}
\DashLine(72.000000,59.820923)(72.000000,54.388888){0.6}
\DashLine(72.000000,54.388888)(76.000000,54.388888){0.6}
\DashLine(76.000000,54.388888)(76.000000,54.980357){0.6}
\DashLine(76.000000,54.980357)(80.000000,54.980357){0.6}
\DashLine(80.000000,54.980357)(80.000000,48.448518){0.6}
\DashLine(80.000000,48.448518)(84.000000,48.448518){0.6}
\DashLine(84.000000,48.448518)(84.000000,42.813597){0.6}
\DashLine(84.000000,42.813597)(88.000000,42.813597){0.6}
\DashLine(88.000000,42.813597)(88.000000,41.349357){0.6}
\DashLine(88.000000,41.349357)(92.000000,41.349357){0.6}
\DashLine(92.000000,41.349357)(92.000000,33.180637){0.6}
\DashLine(92.000000,33.180637)(96.000000,33.180637){0.6}
\DashLine(96.000000,33.180637)(96.000000,35.714438){0.6}
\DashLine(96.000000,35.714438)(100.000000,35.714438){0.6}
\DashLine(100.000000,35.714438)(100.000000,30.079517){0.6}
\DashLine(100.000000,30.079517)(104.000000,30.079517){0.6}
\DashLine(104.000000,30.079517)(104.000000,26.081478){0.6}
\DashLine(104.000000,26.081478)(108.000000,26.081478){0.6}
\DashLine(108.000000,26.081478)(108.000000,0.000000){0.6}
\DashLine(108.000000,0.000000)(112.000000,0.000000){0.6}
\DashLine(112.000000,0.000000)(112.000000,10.813598){0.6}
\DashLine(112.000000,10.813598)(116.000000,10.813598){0.6}
\DashLine(116.000000,10.813598)(120.000000,10.813598){0.6}
\DashLine(120.000000,10.813598)(124.000000,10.813598){0.6}
\DashLine(124.000000,10.813598)(124.000000,0.000000){0.6}
\DashLine(124.000000,0.000000)(128.000000,0.000000){0.6}
\DashLine(128.000000,0.000000)(128.000000,20.446557){0.6}
\DashLine(128.000000,20.446557)(132.000000,20.446557){0.6}
\DashLine(132.000000,20.446557)(132.000000,10.813598){0.6}
\DashLine(132.000000,10.813598)(136.000000,10.813598){0.6}
\DashLine(136.000000,10.813598)(136.000000,0.000000){0.6}
\DashLine(136.000000,0.000000)(140.000000,0.000000){0.6}
\DashLine(140.000000,0.000000)(140.000000,10.813598){0.6}
\DashLine(140.000000,10.813598)(144.000000,10.813598){0.6}
\DashLine(144.000000,10.813598)(144.000000,0.000000){0.6}
\DashLine(144.000000,0.000000)(148.000000,0.000000){0.6}
\DashLine(148.000000,0.000000)(148.000000,10.813598){0.6}
\DashLine(148.000000,10.813598)(152.000000,10.813598){0.6}
\DashLine(152.000000,10.813598)(156.000000,10.813598){0.6}
\DashLine(156.000000,10.813598)(160.000000,10.813598){0.6}
\SetWidth{0.5}
%
%
\Line(0.000000,155.869186)(4.000000,155.869186)
\Line(4.000000,155.869186)(4.000000,147.199423)
\Line(4.000000,147.199423)(8.000000,147.199423)
\Line(8.000000,147.199423)(8.000000,138.448702)
\Line(8.000000,138.448702)(12.000000,138.448702)
\Line(12.000000,138.448702)(12.000000,130.350091)
\Line(12.000000,130.350091)(16.000000,130.350091)
\Line(16.000000,130.350091)(16.000000,122.707451)
\Line(16.000000,122.707451)(20.000000,122.707451)
\Line(20.000000,122.707451)(20.000000,116.305895)
\Line(20.000000,116.305895)(24.000000,116.305895)
\Line(24.000000,116.305895)(24.000000,109.404697)
\Line(24.000000,109.404697)(28.000000,109.404697)
\Line(28.000000,109.404697)(28.000000,104.013330)
\Line(28.000000,104.013330)(32.000000,104.013330)
\Line(32.000000,104.013330)(32.000000,98.092563)
\Line(32.000000,98.092563)(36.000000,98.092563)
\Line(36.000000,98.092563)(36.000000,92.901033)
\Line(36.000000,92.901033)(40.000000,92.901033)
\Line(40.000000,92.901033)(40.000000,88.144138)
\Line(40.000000,88.144138)(44.000000,88.144138)
\Line(44.000000,88.144138)(44.000000,83.468041)
\Line(44.000000,83.468041)(48.000000,83.468041)
\Line(48.000000,83.468041)(48.000000,79.067149)
\Line(48.000000,79.067149)(52.000000,79.067149)
\Line(52.000000,79.067149)(52.000000,73.635113)
\Line(52.000000,73.635113)(56.000000,73.635113)
\Line(56.000000,73.635113)(56.000000,69.158983)
\Line(56.000000,69.158983)(60.000000,69.158983)
\Line(60.000000,69.158983)(60.000000,69.704050)
\Line(60.000000,69.704050)(64.000000,69.704050)
\Line(64.000000,69.704050)(64.000000,60.071090)
\Line(64.000000,60.071090)(68.000000,60.071090)
\Line(68.000000,60.071090)(68.000000,59.526023)
\Line(68.000000,59.526023)(72.000000,59.526023)
\Line(72.000000,59.526023)(72.000000,30.059823)
\Line(72.000000,30.059823)(76.000000,30.059823)
\Line(76.000000,30.059823)(76.000000,41.835080)
\Line(76.000000,41.835080)(80.000000,41.835080)
\Line(80.000000,41.835080)(80.000000,30.059823)
\Line(80.000000,30.059823)(84.000000,30.059823)
\Line(84.000000,30.059823)(84.000000,37.158983)
\Line(84.000000,37.158983)(88.000000,37.158983)
\Line(88.000000,37.158983)(88.000000,24.424903)
\Line(88.000000,24.424903)(92.000000,24.424903)
\Line(92.000000,24.424903)(96.000000,24.424903)
\Line(96.000000,24.424903)(96.000000,14.791943)
\Line(96.000000,14.791943)(100.000000,14.791943)
\Line(100.000000,14.791943)(104.000000,14.791943)
\Line(104.000000,14.791943)(104.000000,0.000000)
\Line(104.000000,0.000000)(108.000000,0.000000)
\Line(108.000000,0.000000)(108.000000,14.791943)
\Line(108.000000,14.791943)(112.000000,14.791943)
\Line(112.000000,14.791943)(116.000000,14.791943)
\Line(116.000000,14.791943)(120.000000,14.791943)
\Line(120.000000,14.791943)(120.000000,0.000000)
\Line(120.000000,0.000000)(124.000000,0.000000)
\Line(124.000000,0.000000)(128.000000,0.000000)
\Line(128.000000,0.000000)(132.000000,0.000000)
\Line(132.000000,0.000000)(136.000000,0.000000)
\Line(136.000000,0.000000)(136.000000,24.424903)
\Line(136.000000,24.424903)(140.000000,24.424903)
\Line(140.000000,24.424903)(140.000000,0.000000)
\Line(140.000000,0.000000)(144.000000,0.000000)
\Line(144.000000,0.000000)(148.000000,0.000000)
\Line(148.000000,0.000000)(152.000000,0.000000)
\Line(152.000000,0.000000)(156.000000,0.000000)
\Line(156.000000,0.000000)(156.000000,14.791943)
\Line(156.000000,14.791943)(160.000000,14.791943)
%
%
\DashLine(0.000000,142.200900)(4.000000,142.200900){2}
\DashLine(4.000000,142.200900)(4.000000,144.477343){2}
\DashLine(4.000000,144.477343)(8.000000,144.477343){2}
\DashLine(8.000000,144.477343)(8.000000,142.243040){2}
\DashLine(8.000000,142.243040)(12.000000,142.243040){2}
\DashLine(12.000000,142.243040)(12.000000,138.226956){2}
\DashLine(12.000000,138.226956)(16.000000,138.226956){2}
\DashLine(16.000000,138.226956)(16.000000,134.451163){2}
\DashLine(16.000000,134.451163)(20.000000,134.451163){2}
\DashLine(20.000000,134.451163)(20.000000,130.257823){2}
\DashLine(20.000000,130.257823)(24.000000,130.257823){2}
\DashLine(24.000000,130.257823)(24.000000,125.388156){2}
\DashLine(24.000000,125.388156)(28.000000,125.388156){2}
\DashLine(28.000000,125.388156)(28.000000,121.011025){2}
\DashLine(28.000000,121.011025)(32.000000,121.011025){2}
\DashLine(32.000000,121.011025)(32.000000,116.522010){2}
\DashLine(32.000000,116.522010)(36.000000,116.522010){2}
\DashLine(36.000000,116.522010)(36.000000,112.754782){2}
\DashLine(36.000000,112.754782)(40.000000,112.754782){2}
\DashLine(40.000000,112.754782)(40.000000,107.515545){2}
\DashLine(40.000000,107.515545)(44.000000,107.515545){2}
\DashLine(44.000000,107.515545)(44.000000,102.214865){2}
\DashLine(44.000000,102.214865)(48.000000,102.214865){2}
\DashLine(48.000000,102.214865)(48.000000,98.172125){2}
\DashLine(48.000000,98.172125)(52.000000,98.172125){2}
\DashLine(52.000000,98.172125)(52.000000,94.174086){2}
\DashLine(52.000000,94.174086)(56.000000,94.174086){2}
\DashLine(56.000000,94.174086)(56.000000,89.507537){2}
\DashLine(56.000000,89.507537)(60.000000,89.507537){2}
\DashLine(60.000000,89.507537)(60.000000,82.541367){2}
\DashLine(60.000000,82.541367)(64.000000,82.541367){2}
\DashLine(64.000000,82.541367)(64.000000,80.527158){2}
\DashLine(64.000000,80.527158)(68.000000,80.527158){2}
\DashLine(68.000000,80.527158)(68.000000,73.969795){2}
\DashLine(68.000000,73.969795)(72.000000,73.969795){2}
\DashLine(72.000000,73.969795)(72.000000,66.595429){2}
\DashLine(72.000000,66.595429)(76.000000,66.595429){2}
\DashLine(76.000000,66.595429)(76.000000,50.975697){2}
\DashLine(76.000000,50.975697)(80.000000,50.975697){2}
\DashLine(80.000000,50.975697)(80.000000,30.597390){2}
\DashLine(80.000000,30.597390)(84.000000,30.597390){2}
\DashLine(84.000000,30.597390)(84.000000,15.329509){2}
\DashLine(84.000000,15.329509)(88.000000,15.329509){2}
\DashLine(88.000000,15.329509)(88.000000,0.000000){2}
\DashLine(88.000000,0.000000)(92.000000,0.000000){2}
\DashLine(92.000000,0.000000)(96.000000,0.000000){2}
\DashLine(96.000000,0.000000)(100.000000,0.000000){2}
\DashLine(100.000000,0.000000)(104.000000,0.000000){2}
\DashLine(104.000000,0.000000)(108.000000,0.000000){2}
\DashLine(108.000000,0.000000)(112.000000,0.000000){2}
\DashLine(112.000000,0.000000)(112.000000,15.329509){2}
\DashLine(112.000000,15.329509)(116.000000,15.329509){2}
\DashLine(116.000000,15.329509)(116.000000,0.000000){2}
\DashLine(116.000000,0.000000)(120.000000,0.000000){2}
\DashLine(120.000000,0.000000)(124.000000,0.000000){2}
\DashLine(124.000000,0.000000)(124.000000,15.329509){2}
\DashLine(124.000000,15.329509)(128.000000,15.329509){2}
\DashLine(128.000000,15.329509)(128.000000,0.000000){2}
\DashLine(128.000000,0.000000)(132.000000,0.000000){2}
\DashLine(132.000000,0.000000)(136.000000,0.000000){2}
\DashLine(136.000000,0.000000)(140.000000,0.000000){2}
\DashLine(140.000000,0.000000)(144.000000,0.000000){2}
\DashLine(144.000000,0.000000)(148.000000,0.000000){2}
\DashLine(148.000000,0.000000)(152.000000,0.000000){2}
\DashLine(152.000000,0.000000)(156.000000,0.000000){2}
\DashLine(156.000000,0.000000)(160.000000,0.000000){2}
%
%
\SetOffset(245,225)
\SetWidth{1.0}\BBox(0,0)(160,160)\SetWidth{0.5}
\LinAxis(0,0)(160,0)(4,5,2.5,0,1)
\LinAxis(0,160)(160,160)(4,5,-2.5,0,1)
\LogAxis(0,0)(0,160)(4,-2.5,3,1)
\LogAxis(160,0)(160,160)(4,2.5,3,1)
\Text(1,-6)[]{0.6}
\Text(41,-6)[]{0.7}
\Text(81,-6)[]{0.8}
\Text(121,-6)[]{0.9}
\Text(161,-6)[]{1.0}
\Text(80,-18)[]{thrust}
\Text(-12,140)[]{$10^1$}
\Text(-12,100)[]{$0$}
\Text(-12,60)[]{$10^{-1}$}
\Text(-12,20)[]{$10^{-2}$}
\rText(-33,80)[][l]{$\frac{1}{\sigma}\frac{d\ \sigma}{d\ thrust}$}
%
%
\DashLine(0.000000,0.000000)(4.000000,0.000000){1.8}
\DashLine(4.000000,0.000000)(8.000000,0.000000){1.8}
\DashLine(8.000000,0.000000)(12.000000,0.000000){1.8}
\DashLine(12.000000,0.000000)(16.000000,0.000000){1.8}
\DashLine(16.000000,0.000000)(20.000000,0.000000){1.8}
\DashLine(20.000000,0.000000)(24.000000,0.000000){1.8}
\DashLine(24.000000,0.000000)(24.000000,19.162751){1.8}
\DashLine(24.000000,19.162751)(28.000000,19.162751){1.8}
\DashLine(28.000000,19.162751)(28.000000,62.617147){1.8}
\DashLine(28.000000,62.617147)(32.000000,62.617147){1.8}
\DashLine(32.000000,62.617147)(32.000000,78.928940){1.8}
\DashLine(32.000000,78.928940)(36.000000,78.928940){1.8}
\DashLine(36.000000,78.928940)(36.000000,87.259974){1.8}
\DashLine(36.000000,87.259974)(40.000000,87.259974){1.8}
\DashLine(40.000000,87.259974)(40.000000,91.944508){1.8}
\DashLine(40.000000,91.944508)(44.000000,91.944508){1.8}
\DashLine(44.000000,91.944508)(44.000000,95.459207){1.8}
\DashLine(44.000000,95.459207)(48.000000,95.459207){1.8}
\DashLine(48.000000,95.459207)(48.000000,98.669402){1.8}
\DashLine(48.000000,98.669402)(52.000000,98.669402){1.8}
\DashLine(52.000000,98.669402)(52.000000,102.301025){1.8}
\DashLine(52.000000,102.301025)(56.000000,102.301025){1.8}
\DashLine(56.000000,102.301025)(56.000000,104.064497){1.8}
\DashLine(56.000000,104.064497)(60.000000,104.064497){1.8}
\DashLine(60.000000,104.064497)(60.000000,106.573103){1.8}
\DashLine(60.000000,106.573103)(64.000000,106.573103){1.8}
\DashLine(64.000000,106.573103)(64.000000,110.494797){1.8}
\DashLine(64.000000,110.494797)(68.000000,110.494797){1.8}
\DashLine(68.000000,110.494797)(68.000000,111.479695){1.8}
\DashLine(68.000000,111.479695)(72.000000,111.479695){1.8}
\DashLine(72.000000,111.479695)(72.000000,113.063756){1.8}
\DashLine(72.000000,113.063756)(76.000000,113.063756){1.8}
\DashLine(76.000000,113.063756)(76.000000,116.621980){1.8}
\DashLine(76.000000,116.621980)(80.000000,116.621980){1.8}
\DashLine(80.000000,116.621980)(80.000000,117.670651){1.8}
\DashLine(80.000000,117.670651)(84.000000,117.670651){1.8}
\DashLine(84.000000,117.670651)(84.000000,118.131401){1.8}
\DashLine(84.000000,118.131401)(88.000000,118.131401){1.8}
\DashLine(88.000000,118.131401)(88.000000,120.775930){1.8}
\DashLine(88.000000,120.775930)(92.000000,120.775930){1.8}
\DashLine(92.000000,120.775930)(92.000000,123.392190){1.8}
\DashLine(92.000000,123.392190)(96.000000,123.392190){1.8}
\DashLine(96.000000,123.392190)(96.000000,124.142284){1.8}
\DashLine(96.000000,124.142284)(100.000000,124.142284){1.8}
\DashLine(100.000000,124.142284)(100.000000,126.419638){1.8}
\DashLine(100.000000,126.419638)(104.000000,126.419638){1.8}
\DashLine(104.000000,126.419638)(104.000000,127.949259){1.8}
\DashLine(104.000000,127.949259)(108.000000,127.949259){1.8}
\DashLine(108.000000,127.949259)(108.000000,129.890511){1.8}
\DashLine(108.000000,129.890511)(112.000000,129.890511){1.8}
\DashLine(112.000000,129.890511)(112.000000,131.339235){1.8}
\DashLine(112.000000,131.339235)(116.000000,131.339235){1.8}
\DashLine(116.000000,131.339235)(116.000000,133.422242){1.8}
\DashLine(116.000000,133.422242)(120.000000,133.422242){1.8}
\DashLine(120.000000,133.422242)(120.000000,135.037023){1.8}
\DashLine(120.000000,135.037023)(124.000000,135.037023){1.8}
\DashLine(124.000000,135.037023)(124.000000,137.189027){1.8}
\DashLine(124.000000,137.189027)(128.000000,137.189027){1.8}
\DashLine(128.000000,137.189027)(128.000000,138.260720){1.8}
\DashLine(128.000000,138.260720)(132.000000,138.260720){1.8}
\DashLine(132.000000,138.260720)(132.000000,139.605260){1.8}
\DashLine(132.000000,139.605260)(136.000000,139.605260){1.8}
\DashLine(136.000000,139.605260)(136.000000,139.870792){1.8}
\DashLine(136.000000,139.870792)(140.000000,139.870792){1.8}
\DashLine(140.000000,139.870792)(140.000000,63.177578){1.8}
\DashLine(140.000000,63.177578)(144.000000,0.0){1.8}
%
%
\Line(0.000000,0.000000)(4.000000,0.000000)
\Line(4.000000,0.000000)(8.000000,0.000000)
\Line(8.000000,0.000000)(12.000000,0.000000)
\Line(12.000000,0.000000)(16.000000,0.000000)
\Line(16.000000,0.000000)(20.000000,0.000000)
\Line(20.000000,0.000000)(24.000000,0.000000)
\Line(24.000000,0.000000)(24.000000,21.656879)
\Line(24.000000,21.656879)(28.000000,21.656879)
\Line(28.000000,21.656879)(28.000000,60.145337)
\Line(28.000000,60.145337)(32.000000,60.145337)
\Line(32.000000,60.145337)(32.000000,76.499595)
\Line(32.000000,76.499595)(36.000000,76.499595)
\Line(36.000000,76.499595)(36.000000,78.366815)
\Line(36.000000,78.366815)(40.000000,78.366815)
\Line(40.000000,78.366815)(40.000000,83.989219)
\Line(40.000000,83.989219)(44.000000,83.989219)
\Line(44.000000,83.989219)(44.000000,90.682455)
\Line(44.000000,90.682455)(48.000000,90.682455)
\Line(48.000000,90.682455)(48.000000,94.128812)
\Line(48.000000,94.128812)(52.000000,94.128812)
\Line(52.000000,94.128812)(52.000000,97.525275)
\Line(52.000000,97.525275)(56.000000,97.525275)
\Line(56.000000,97.525275)(56.000000,99.204338)
\Line(56.000000,99.204338)(60.000000,99.204338)
\Line(60.000000,99.204338)(60.000000,102.198455)
\Line(60.000000,102.198455)(64.000000,102.198455)
\Line(64.000000,102.198455)(64.000000,104.309920)
\Line(64.000000,104.309920)(68.000000,104.309920)
\Line(68.000000,104.309920)(68.000000,106.102900)
\Line(68.000000,106.102900)(72.000000,106.102900)
\Line(72.000000,106.102900)(72.000000,108.545422)
\Line(72.000000,108.545422)(76.000000,108.545422)
\Line(76.000000,108.545422)(76.000000,110.232972)
\Line(76.000000,110.232972)(80.000000,110.232972)
\Line(80.000000,110.232972)(80.000000,112.484032)
\Line(80.000000,112.484032)(84.000000,112.484032)
\Line(84.000000,112.484032)(84.000000,114.267699)
\Line(84.000000,114.267699)(88.000000,114.267699)
\Line(88.000000,114.267699)(88.000000,116.597881)
\Line(88.000000,116.597881)(92.000000,116.597881)
\Line(92.000000,116.597881)(92.000000,118.455109)
\Line(92.000000,118.455109)(96.000000,118.455109)
\Line(96.000000,118.455109)(96.000000,121.000371)
\Line(96.000000,121.000371)(100.000000,121.000371)
\Line(100.000000,121.000371)(100.000000,123.144449)
\Line(100.000000,123.144449)(104.000000,123.144449)
\Line(104.000000,123.144449)(104.000000,125.037666)
\Line(104.000000,125.037666)(108.000000,125.037666)
\Line(108.000000,125.037666)(108.000000,127.707997)
\Line(108.000000,127.707997)(112.000000,127.707997)
\Line(112.000000,127.707997)(112.000000,130.038986)
\Line(112.000000,130.038986)(116.000000,130.038986)
\Line(116.000000,130.038986)(116.000000,132.840741)
\Line(116.000000,132.840741)(120.000000,132.840741)
\Line(120.000000,132.840741)(120.000000,135.290386)
\Line(120.000000,135.290386)(124.000000,135.290386)
\Line(124.000000,135.290386)(124.000000,137.569291)
\Line(124.000000,137.569291)(128.000000,137.569291)
\Line(128.000000,137.569291)(128.000000,141.121769)
\Line(128.000000,141.121769)(132.000000,141.121769)
\Line(132.000000,141.121769)(132.000000,144.522339)
\Line(132.000000,144.522339)(136.000000,144.522339)
\Line(136.000000,144.522339)(136.000000,147.687135)
\Line(136.000000,147.687135)(140.000000,147.687135)
\Line(140.000000,147.687135)(140.000000,49.026749)
\Line(140.000000,49.026749)(144.000000,0.0)
%
%
\SetOffset(40,25)
\SetWidth{1.0}\BBox(0,0)(160,160)\SetWidth{0.5}
\LinAxis(0,0)(160,0)(10,1,2.5,0,1)
\LinAxis(0,160)(160,160)(10,1,-2.5,0,1)
\LinAxis(0,0)(0,160)(7,2,-2.5,0,1)
\LinAxis(160,0)(160,160)(7,2,2.5,0,1)
\Text(1,-6)[]{0}
\Text(33,-6)[]{0.2}
\Text(65,-6)[]{0.4}
\Text(97,-6)[]{0.6}
\Text(129,-6)[]{0.8}
\Text(161,-6)[]{1.0}
\Text(80,-18)[]{$x_q$}
\Text(-8,137)[]{6}
\Text(-8,114)[]{5}
\Text(-8,91)[]{4}
\Text(-8,69)[]{3}
\Text(-8,46)[]{2}
\Text(-8,23)[]{1}
\rText(-26,80)[][l]{$\frac{1}{\sigma}\frac{d\ \sigma}{d\ x_q}$}
%
%
\Line(0.000000,0.000000)(3.200000,0.000000)
\Line(3.200000,0.000000)(6.400000,0.000000)
\Line(6.400000,0.000000)(9.600000,0.000000)
\Line(9.600000,0.000000)(12.800000,0.000000)
\Line(12.800000,0.000000)(16.000000,0.000000)
\Line(16.000000,0.000000)(16.000000,0.139150)
\Line(16.000000,0.139150)(19.200000,0.139150)
\Line(19.200000,0.139150)(19.200000,0.397571)
\Line(19.200000,0.397571)(22.400000,0.397571)
\Line(22.400000,0.397571)(22.400000,0.815020)
\Line(22.400000,0.815020)(25.600000,0.815020)
\Line(25.600000,0.815020)(25.600000,1.749311)
\Line(25.600000,1.749311)(28.800000,1.749311)
\Line(28.800000,1.749311)(28.800000,2.266153)
\Line(28.800000,2.266153)(32.000000,2.266153)
\Line(32.000000,2.266153)(32.000000,2.464939)
\Line(32.000000,2.464939)(35.200000,2.464939)
\Line(35.200000,2.464939)(35.200000,3.598016)
\Line(35.200000,3.598016)(38.400000,3.598016)
\Line(38.400000,3.598016)(38.400000,3.796800)
\Line(38.400000,3.796800)(41.600000,3.796800)
\Line(41.600000,3.796800)(41.600000,5.148542)
\Line(41.600000,5.148542)(44.800000,5.148542)
\Line(44.800000,5.148542)(44.800000,5.247934)
\Line(44.800000,5.247934)(48.000000,5.247934)
\Line(48.000000,5.247934)(48.000000,5.585870)
\Line(48.000000,5.585870)(51.200000,5.585870)
\Line(51.200000,5.585870)(51.200000,6.003319)
\Line(51.200000,6.003319)(54.400000,6.003319)
\Line(54.400000,6.003319)(54.400000,6.897853)
\Line(54.400000,6.897853)(57.600000,6.897853)
\Line(57.600000,6.897853)(57.600000,6.937609)
\Line(57.600000,6.937609)(60.800000,6.937609)
\Line(60.800000,6.937609)(60.800000,7.374937)
\Line(60.800000,7.374937)(64.000000,7.374937)
\Line(64.000000,7.374937)(64.000000,8.329106)
\Line(64.000000,8.329106)(67.200000,8.329106)
\Line(67.200000,8.329106)(67.200000,9.601333)
\Line(67.200000,9.601333)(70.400000,9.601333)
\Line(70.400000,9.601333)(70.400000,9.899513)
\Line(70.400000,9.899513)(73.600000,9.899513)
\Line(73.600000,9.899513)(73.600000,11.291010)
\Line(73.600000,11.291010)(76.800000,11.291010)
\Line(76.800000,11.291010)(76.800000,11.489795)
\Line(76.800000,11.489795)(80.000000,11.489795)
\Line(80.000000,11.489795)(80.000000,12.225301)
\Line(80.000000,12.225301)(83.200000,12.225301)
\Line(83.200000,12.225301)(83.200000,14.729996)
\Line(83.200000,14.729996)(86.400000,14.729996)
\Line(86.400000,14.729996)(86.400000,14.252912)
\Line(86.400000,14.252912)(89.600000,14.252912)
\Line(89.600000,14.252912)(89.600000,16.538944)
\Line(89.600000,16.538944)(92.800000,16.538944)
\Line(92.800000,16.538944)(92.800000,17.135300)
\Line(92.800000,17.135300)(96.000000,17.135300)
\Line(96.000000,17.135300)(96.000000,18.964123)
\Line(96.000000,18.964123)(99.200000,18.964123)
\Line(99.200000,18.964123)(99.200000,19.401451)
\Line(99.200000,19.401451)(102.400000,19.401451)
\Line(102.400000,19.401451)(102.400000,22.184448)
\Line(102.400000,22.184448)(105.600000,22.184448)
\Line(105.600000,22.184448)(105.600000,25.007199)
\Line(105.600000,25.007199)(108.800000,25.007199)
\Line(108.800000,25.007199)(108.800000,26.100527)
\Line(108.800000,26.100527)(112.000000,26.100527)
\Line(112.000000,26.100527)(112.000000,29.062424)
\Line(112.000000,29.062424)(115.200000,29.062424)
\Line(115.200000,29.062424)(115.200000,30.771978)
\Line(115.200000,30.771978)(118.400000,30.771978)
\Line(118.400000,30.771978)(118.400000,33.654355)
\Line(118.400000,33.654355)(121.600000,33.654355)
\Line(121.600000,33.654355)(121.600000,39.021576)
\Line(121.600000,39.021576)(124.800000,39.021576)
\Line(124.800000,39.021576)(124.800000,43.295452)
\Line(124.800000,43.295452)(128.000000,43.295452)
\Line(128.000000,43.295452)(128.000000,48.404252)
\Line(128.000000,48.404252)(131.200000,48.404252)
\Line(131.200000,48.404252)(131.200000,57.429096)
\Line(131.200000,57.429096)(134.400000,57.429096)
\Line(134.400000,57.429096)(134.400000,67.924957)
\Line(134.400000,67.924957)(137.600000,67.924957)
\Line(137.600000,67.924957)(137.600000,74.842698)
\Line(137.600000,74.842698)(140.800000,74.842698)
\Line(140.800000,74.842698)(140.800000,93.608028)
\Line(140.800000,93.608028)(144.000000,93.608028)
\Line(144.000000,93.608028)(144.000000,108.079616)
\Line(144.000000,108.079616)(147.200000,108.079616)
\Line(147.200000,108.079616)(147.200000,137.817895)
\Line(147.200000,137.817895)(150.400000,137.817895)
\Line(150.400000,137.817895)(150.400000,82.654948)
\Line(150.400000,82.654948)(153.600000,82.654948)
\Line(153.600000,82.654948)(153.600000,0.536720)
\Line(153.600000,0.536720)(156.800000,0.536720)
\Line(156.800000,0.536720)(156.800000,0.119271)
\Line(156.800000,0.119271)(160.000000,0.119271)
%
%
\DashLine(0.000000,0.000000)(3.200000,0.000000){1.5}
\DashLine(3.200000,0.000000)(6.400000,0.000000){1.5}
\DashLine(6.400000,0.000000)(9.600000,0.000000){1.5}
\DashLine(9.600000,0.000000)(12.800000,0.000000){1.5}
\DashLine(12.800000,0.000000)(16.000000,0.000000){1.5}
\DashLine(16.000000,0.000000)(16.000000,0.785228){1.5}
\DashLine(16.000000,0.785228)(19.200000,0.785228){1.5}
\DashLine(19.200000,0.785228)(19.200000,3.409541){1.5}
\DashLine(19.200000,3.409541)(22.400000,3.409541){1.5}
\DashLine(22.400000,3.409541)(22.400000,7.170368){1.5}
\DashLine(22.400000,7.170368)(25.600000,7.170368){1.5}
\DashLine(25.600000,7.170368)(25.600000,10.827876){1.5}
\DashLine(25.600000,10.827876)(28.800000,10.827876){1.5}
\DashLine(28.800000,10.827876)(28.800000,13.555509){1.5}
\DashLine(28.800000,13.555509)(32.000000,13.555509){1.5}
\DashLine(32.000000,13.555509)(32.000000,16.655092){1.5}
\DashLine(32.000000,16.655092)(35.200000,16.655092){1.5}
\DashLine(35.200000,16.655092)(35.200000,16.386461){1.5}
\DashLine(35.200000,16.386461)(38.400000,16.386461){1.5}
\DashLine(38.400000,16.386461)(38.400000,19.403390){1.5}
\DashLine(38.400000,19.403390)(41.600000,19.403390){1.5}
\DashLine(41.600000,19.403390)(41.600000,18.432188){1.5}
\DashLine(41.600000,18.432188)(44.800000,18.432188){1.5}
\DashLine(44.800000,18.432188)(44.800000,19.382725){1.5}
\DashLine(44.800000,19.382725)(48.000000,19.382725){1.5}
\DashLine(48.000000,19.382725)(48.000000,20.477911){1.5}
\DashLine(48.000000,20.477911)(51.200000,20.477911){1.5}
\DashLine(51.200000,20.477911)(51.200000,20.849860){1.5}
\DashLine(51.200000,20.849860)(54.400000,20.849860){1.5}
\DashLine(54.400000,20.849860)(54.400000,19.837330){1.5}
\DashLine(54.400000,19.837330)(57.600000,19.837330){1.5}
\DashLine(57.600000,19.837330)(57.600000,21.180483){1.5}
\DashLine(57.600000,21.180483)(60.800000,21.180483){1.5}
\DashLine(60.800000,21.180483)(60.800000,23.122880){1.5}
\DashLine(60.800000,23.122880)(64.000000,23.122880){1.5}
\DashLine(64.000000,23.122880)(64.000000,22.110356){1.5}
\DashLine(64.000000,22.110356)(67.200000,22.110356){1.5}
\DashLine(67.200000,22.110356)(67.200000,24.032091){1.5}
\DashLine(67.200000,24.032091)(70.400000,24.032091){1.5}
\DashLine(70.400000,24.032091)(70.400000,25.333920){1.5}
\DashLine(70.400000,25.333920)(73.600000,25.333920){1.5}
\DashLine(73.600000,25.333920)(73.600000,24.837990){1.5}
\DashLine(73.600000,24.837990)(76.800000,24.837990){1.5}
\DashLine(76.800000,24.837990)(76.800000,23.804801){1.5}
\DashLine(76.800000,23.804801)(80.000000,23.804801){1.5}
\DashLine(80.000000,23.804801)(80.000000,26.470446){1.5}
\DashLine(80.000000,26.470446)(83.200000,26.470446){1.5}
\DashLine(83.200000,26.470446)(83.200000,25.643885){1.5}
\DashLine(83.200000,25.643885)(86.400000,25.643885){1.5}
\DashLine(86.400000,25.643885)(86.400000,27.503635){1.5}
\DashLine(86.400000,27.503635)(89.600000,27.503635){1.5}
\DashLine(89.600000,27.503635)(89.600000,26.945714){1.5}
\DashLine(89.600000,26.945714)(92.800000,26.945714){1.5}
\DashLine(92.800000,26.945714)(92.800000,27.916914){1.5}
\DashLine(92.800000,27.916914)(96.000000,27.916914){1.5}
\DashLine(96.000000,27.916914)(96.000000,27.586286){1.5}
\DashLine(96.000000,27.586286)(99.200000,27.586286){1.5}
\DashLine(99.200000,27.586286)(99.200000,27.668937){1.5}
\DashLine(99.200000,27.668937)(102.400000,27.668937){1.5}
\DashLine(102.400000,27.668937)(102.400000,28.681485){1.5}
\DashLine(102.400000,28.681485)(105.600000,28.681485){1.5}
\DashLine(105.600000,28.681485)(105.600000,29.941965){1.5}
\DashLine(105.600000,29.941965)(108.800000,29.941965){1.5}
\DashLine(108.800000,29.941965)(108.800000,30.210607){1.5}
\DashLine(108.800000,30.210607)(112.000000,30.210607){1.5}
\DashLine(112.000000,30.210607)(112.000000,30.954492){1.5}
\DashLine(112.000000,30.954492)(115.200000,30.954492){1.5}
\DashLine(115.200000,30.954492)(115.200000,33.165531){1.5}
\DashLine(115.200000,33.165531)(118.400000,33.165531){1.5}
\DashLine(118.400000,33.165531)(118.400000,33.062218){1.5}
\DashLine(118.400000,33.062218)(121.600000,33.062218){1.5}
\DashLine(121.600000,33.062218)(121.600000,34.839314){1.5}
\DashLine(121.600000,34.839314)(124.800000,34.839314){1.5}
\DashLine(124.800000,34.839314)(124.800000,37.071018){1.5}
\DashLine(124.800000,37.071018)(128.000000,37.071018){1.5}
\DashLine(128.000000,37.071018)(128.000000,37.215658){1.5}
\DashLine(128.000000,37.215658)(131.200000,37.215658){1.5}
\DashLine(131.200000,37.215658)(131.200000,39.777987){1.5}
\DashLine(131.200000,39.777987)(134.400000,39.777987){1.5}
\DashLine(134.400000,39.777987)(134.400000,41.803041){1.5}
\DashLine(134.400000,41.803041)(137.600000,41.803041){1.5}
\DashLine(137.600000,41.803041)(137.600000,44.158720){1.5}
\DashLine(137.600000,44.158720)(140.800000,44.158720){1.5}
\DashLine(140.800000,44.158720)(140.800000,48.498148){1.5}
\DashLine(140.800000,48.498148)(144.000000,48.498148){1.5}
\DashLine(144.000000,48.498148)(144.000000,51.081118){1.5}
\DashLine(144.000000,51.081118)(147.200000,51.081118){1.5}
\DashLine(147.200000,51.081118)(147.200000,54.883292){1.5}
\DashLine(147.200000,54.883292)(150.400000,54.883292){1.5}
\DashLine(150.400000,54.883292)(150.400000,26.139818){1.5}
\DashLine(150.400000,26.139818)(153.600000,26.139818){1.5}
\DashLine(153.600000,26.139818)(153.600000,0.000000){1.5}
\DashLine(153.600000,0.000000)(156.800000,0.000000){1.5}
\DashLine(156.800000,0.000000)(156.800000,0.041328){1.5}
\DashLine(156.800000,0.041328)(160.000000,0.041328){1.5}
%
%
\SetOffset(245,25)
\SetWidth{1.0}\BBox(0,0)(160,160)\SetWidth{0.5}
\LinAxis(0,0)(160,0)(10,1,2.5,0,1)
\LinAxis(0,160)(160,160)(10,1,-2.5,0,1)
\LinAxis(0,0)(0,160)(7,2,-2.5,0,1)
\LinAxis(160,0)(160,160)(7,2,2.5,0,1)
\Text(1,-6)[]{0}
\Text(33,-6)[]{0.2}
\Text(65,-6)[]{0.4}
\Text(97,-6)[]{0.6}
\Text(129,-6)[]{0.8}
\Text(161,-6)[]{1.0}
\Text(80,-18)[]{$x_g$}
\Text(-8,137)[]{6}
\Text(-8,114)[]{5}
\Text(-8,91)[]{4}
\Text(-8,69)[]{3}
\Text(-8,46)[]{2}
\Text(-8,23)[]{1}
\rText(-26,80)[][l]{$\frac{1}{\sigma}\frac{d\ \sigma}{d\ x_g}$}
%
%
\Line(0.000000,0.000000)(3.200000,0.000000)
\Line(3.200000,0.000000)(6.400000,0.000000)
\Line(6.400000,0.000000)(9.600000,0.000000)
\Line(9.600000,0.000000)(9.600000,0.019879)
\Line(9.600000,0.019879)(12.800000,0.019879)
\Line(12.800000,0.019879)(12.800000,0.000000)
\Line(12.800000,0.000000)(16.000000,0.000000)
\Line(16.000000,0.000000)(16.000000,8.547771)
\Line(16.000000,8.547771)(19.200000,8.547771)
\Line(19.200000,8.547771)(19.200000,25.166217)
\Line(19.200000,25.166217)(22.400000,25.166217)
\Line(22.400000,25.166217)(22.400000,32.362262)
\Line(22.400000,32.362262)(25.600000,32.362262)
\Line(25.600000,32.362262)(25.600000,40.134765)
\Line(25.600000,40.134765)(28.800000,40.134765)
\Line(28.800000,40.134765)(28.800000,43.494241)
\Line(28.800000,43.494241)(32.000000,43.494241)
\Line(32.000000,43.494241)(32.000000,43.076801)
\Line(32.000000,43.076801)(35.200000,43.076801)
\Line(35.200000,43.076801)(35.200000,45.104411)
\Line(35.200000,45.104411)(38.400000,45.104411)
\Line(38.400000,45.104411)(38.400000,44.309256)
\Line(38.400000,44.309256)(41.600000,44.309256)
\Line(41.600000,44.309256)(41.600000,43.215931)
\Line(41.600000,43.215931)(44.800000,43.215931)
\Line(44.800000,43.215931)(44.800000,40.969669)
\Line(44.800000,40.969669)(48.000000,40.969669)
\Line(48.000000,40.969669)(48.000000,39.538400)
\Line(48.000000,39.538400)(51.200000,39.538400)
\Line(51.200000,39.538400)(51.200000,37.968001)
\Line(51.200000,37.968001)(54.400000,37.968001)
\Line(54.400000,37.968001)(54.400000,36.099428)
\Line(54.400000,36.099428)(57.600000,36.099428)
\Line(57.600000,36.099428)(57.600000,36.377713)
\Line(57.600000,36.377713)(60.800000,36.377713)
\Line(60.800000,36.377713)(60.800000,34.290469)
\Line(60.800000,34.290469)(64.000000,34.290469)
\Line(64.000000,34.290469)(64.000000,32.898971)
\Line(64.000000,32.898971)(67.200000,32.898971)
\Line(67.200000,32.898971)(67.200000,33.515222)
\Line(67.200000,33.515222)(70.400000,33.515222)
\Line(70.400000,33.515222)(70.400000,30.215383)
\Line(70.400000,30.215383)(73.600000,30.215383)
\Line(73.600000,30.215383)(73.600000,29.817805)
\Line(73.600000,29.817805)(76.800000,29.817805)
\Line(76.800000,29.817805)(76.800000,28.943154)
\Line(76.800000,28.943154)(80.000000,28.943154)
\Line(80.000000,28.943154)(80.000000,28.128136)
\Line(80.000000,28.128136)(83.200000,28.128136)
\Line(83.200000,28.128136)(83.200000,27.591406)
\Line(83.200000,27.591406)(86.400000,27.591406)
\Line(86.400000,27.591406)(86.400000,26.140273)
\Line(86.400000,26.140273)(89.600000,26.140273)
\Line(89.600000,26.140273)(89.600000,25.504161)
\Line(89.600000,25.504161)(92.800000,25.504161)
\Line(92.800000,25.504161)(92.800000,24.689143)
\Line(92.800000,24.689143)(96.000000,24.689143)
\Line(96.000000,24.689143)(96.000000,23.456685)
\Line(96.000000,23.456685)(99.200000,23.456685)
\Line(99.200000,23.456685)(99.200000,23.754858)
\Line(99.200000,23.754858)(102.400000,23.754858)
\Line(102.400000,23.754858)(102.400000,23.118744)
\Line(102.400000,23.118744)(105.600000,23.118744)
\Line(105.600000,23.118744)(105.600000,20.733314)
\Line(105.600000,20.733314)(108.800000,20.733314)
\Line(108.800000,20.733314)(108.800000,20.991735)
\Line(108.800000,20.991735)(112.000000,20.991735)
\Line(112.000000,20.991735)(112.000000,19.739388)
\Line(112.000000,19.739388)(115.200000,19.739388)
\Line(115.200000,19.739388)(115.200000,19.480967)
\Line(115.200000,19.480967)(118.400000,19.480967)
\Line(118.400000,19.480967)(118.400000,19.202667)
\Line(118.400000,19.202667)(121.600000,19.202667)
\Line(121.600000,19.202667)(121.600000,17.930441)
\Line(121.600000,17.930441)(124.800000,17.930441)
\Line(124.800000,17.930441)(124.800000,16.280522)
\Line(124.800000,16.280522)(128.000000,16.280522)
\Line(128.000000,16.280522)(128.000000,16.459428)
\Line(128.000000,16.459428)(131.200000,16.459428)
\Line(131.200000,16.459428)(131.200000,15.028174)
\Line(131.200000,15.028174)(134.400000,15.028174)
\Line(134.400000,15.028174)(134.400000,14.372183)
\Line(134.400000,14.372183)(137.600000,14.372183)
\Line(137.600000,14.372183)(137.600000,14.789632)
\Line(137.600000,14.789632)(140.800000,14.789632)
\Line(140.800000,14.789632)(140.800000,12.662629)
\Line(140.800000,12.662629)(144.000000,12.662629)
\Line(144.000000,12.662629)(144.000000,12.284934)
\Line(144.000000,12.284934)(147.200000,12.284934)
\Line(147.200000,12.284934)(147.200000,10.098297)
\Line(147.200000,10.098297)(150.400000,10.098297)
\Line(150.400000,10.098297)(150.400000,4.353399)
\Line(150.400000,4.353399)(153.600000,4.353399)
\Line(153.600000,4.353399)(153.600000,0.000000)
\Line(153.600000,0.000000)(156.800000,0.000000)
\Line(156.800000,0.000000)(160.000000,0.000000)
%
%
\DashLine(0.000000,0.000000)(3.200000,0.000000){1.5}
\DashLine(3.200000,0.000000)(6.400000,0.000000){1.5}
\DashLine(6.400000,0.000000)(9.600000,0.000000){1.5}
\DashLine(9.600000,0.000000)(12.800000,0.000000){1.5}
\DashLine(12.800000,0.000000)(16.000000,0.000000){1.5}
\DashLine(16.000000,0.000000)(16.000000,0.123983){1.5}
\DashLine(16.000000,0.123983)(19.200000,0.123983){1.5}
\DashLine(19.200000,0.123983)(19.200000,0.413278){1.5}
\DashLine(19.200000,0.413278)(22.400000,0.413278){1.5}
\DashLine(22.400000,0.413278)(22.400000,1.198505){1.5}
\DashLine(22.400000,1.198505)(25.600000,1.198505){1.5}
\DashLine(25.600000,1.198505)(25.600000,1.797758){1.5}
\DashLine(25.600000,1.797758)(28.800000,1.797758){1.5}
\DashLine(28.800000,1.797758)(28.800000,1.880414){1.5}
\DashLine(28.800000,1.880414)(32.000000,1.880414){1.5}
\DashLine(32.000000,1.880414)(32.000000,3.244231){1.5}
\DashLine(32.000000,3.244231)(35.200000,3.244231){1.5}
\DashLine(35.200000,3.244231)(35.200000,3.802155){1.5}
\DashLine(35.200000,3.802155)(38.400000,3.802155){1.5}
\DashLine(38.400000,3.802155)(38.400000,4.504727){1.5}
\DashLine(38.400000,4.504727)(41.600000,4.504727){1.5}
\DashLine(41.600000,4.504727)(41.600000,4.608046){1.5}
\DashLine(41.600000,4.608046)(44.800000,4.608046){1.5}
\DashLine(44.800000,4.608046)(44.800000,5.806551){1.5}
\DashLine(44.800000,5.806551)(48.000000,5.806551){1.5}
\DashLine(48.000000,5.806551)(48.000000,6.591781){1.5}
\DashLine(48.000000,6.591781)(51.200000,6.591781){1.5}
\DashLine(51.200000,6.591781)(51.200000,6.571116){1.5}
\DashLine(51.200000,6.571116)(54.400000,6.571116){1.5}
\DashLine(54.400000,6.571116)(54.400000,7.315015){1.5}
\DashLine(54.400000,7.315015)(57.600000,7.315015){1.5}
\DashLine(57.600000,7.315015)(57.600000,9.009454){1.5}
\DashLine(57.600000,9.009454)(60.800000,9.009454){1.5}
\DashLine(60.800000,9.009454)(60.800000,8.286217){1.5}
\DashLine(60.800000,8.286217)(64.000000,8.286217){1.5}
\DashLine(64.000000,8.286217)(64.000000,9.340076){1.5}
\DashLine(64.000000,9.340076)(67.200000,9.340076){1.5}
\DashLine(67.200000,9.340076)(67.200000,10.683230){1.5}
\DashLine(67.200000,10.683230)(70.400000,10.683230){1.5}
\DashLine(70.400000,10.683230)(70.400000,10.476590){1.5}
\DashLine(70.400000,10.476590)(73.600000,10.476590){1.5}
\DashLine(73.600000,10.476590)(73.600000,12.790947){1.5}
\DashLine(73.600000,12.790947)(76.800000,12.790947){1.5}
\DashLine(76.800000,12.790947)(76.800000,12.418995){1.5}
\DashLine(76.800000,12.418995)(80.000000,12.418995){1.5}
\DashLine(80.000000,12.418995)(80.000000,15.063972){1.5}
\DashLine(80.000000,15.063972)(83.200000,15.063972){1.5}
\DashLine(83.200000,15.063972)(83.200000,14.981316){1.5}
\DashLine(83.200000,14.981316)(86.400000,14.981316){1.5}
\DashLine(86.400000,14.981316)(86.400000,16.820405){1.5}
\DashLine(86.400000,16.820405)(89.600000,16.820405){1.5}
\DashLine(89.600000,16.820405)(89.600000,18.060236){1.5}
\DashLine(89.600000,18.060236)(92.800000,18.060236){1.5}
\DashLine(92.800000,18.060236)(92.800000,18.824799){1.5}
\DashLine(92.800000,18.824799)(96.000000,18.824799){1.5}
\DashLine(96.000000,18.824799)(96.000000,19.672018){1.5}
\DashLine(96.000000,19.672018)(99.200000,19.672018){1.5}
\DashLine(99.200000,19.672018)(99.200000,24.156091){1.5}
\DashLine(99.200000,24.156091)(102.400000,24.156091){1.5}
\DashLine(102.400000,24.156091)(102.400000,24.218080){1.5}
\DashLine(102.400000,24.218080)(105.600000,24.218080){1.5}
\DashLine(105.600000,24.218080)(105.600000,28.226880){1.5}
\DashLine(105.600000,28.226880)(108.800000,28.226880){1.5}
\DashLine(108.800000,28.226880)(108.800000,28.867452){1.5}
\DashLine(108.800000,28.867452)(112.000000,28.867452){1.5}
\DashLine(112.000000,28.867452)(112.000000,31.409097){1.5}
\DashLine(112.000000,31.409097)(115.200000,31.409097){1.5}
\DashLine(115.200000,31.409097)(115.200000,33.475497){1.5}
\DashLine(115.200000,33.475497)(118.400000,33.475497){1.5}
\DashLine(118.400000,33.475497)(118.400000,38.124869){1.5}
\DashLine(118.400000,38.124869)(121.600000,38.124869){1.5}
\DashLine(121.600000,38.124869)(121.600000,41.534399){1.5}
\DashLine(121.600000,41.534399)(124.800000,41.534399){1.5}
\DashLine(124.800000,41.534399)(124.800000,45.419222){1.5}
\DashLine(124.800000,45.419222)(128.000000,45.419222){1.5}
\DashLine(128.000000,45.419222)(128.000000,52.258971){1.5}
\DashLine(128.000000,52.258971)(131.200000,52.258971){1.5}
\DashLine(131.200000,52.258971)(131.200000,59.388003){1.5}
\DashLine(131.200000,59.388003)(134.400000,59.388003){1.5}
\DashLine(134.400000,59.388003)(134.400000,69.471975){1.5}
\DashLine(134.400000,69.471975)(137.600000,69.471975){1.5}
\DashLine(137.600000,69.471975)(137.600000,77.861530){1.5}
\DashLine(137.600000,77.861530)(140.800000,77.861530){1.5}
\DashLine(140.800000,77.861530)(140.800000,88.565418){1.5}
\DashLine(140.800000,88.565418)(144.000000,88.565418){1.5}
\DashLine(144.000000,88.565418)(144.000000,103.712049){1.5}
\DashLine(144.000000,103.712049)(147.200000,103.712049){1.5}
\DashLine(147.200000,103.712049)(147.200000,115.614450){1.5}
\DashLine(147.200000,115.614450)(150.400000,115.614450){1.5}
\DashLine(150.400000,115.614450)(150.400000,55.544526){1.5}
\DashLine(150.400000,55.544526)(153.600000,55.544526){1.5}
\DashLine(153.600000,55.544526)(153.600000,0.599253){1.5}
\DashLine(153.600000,0.599253)(156.800000,0.599253){1.5}
\DashLine(156.800000,0.599253)(156.800000,0.123983){1.5}
\DashLine(156.800000,0.123983)(160.000000,0.123983){1.5}
\end{picture}
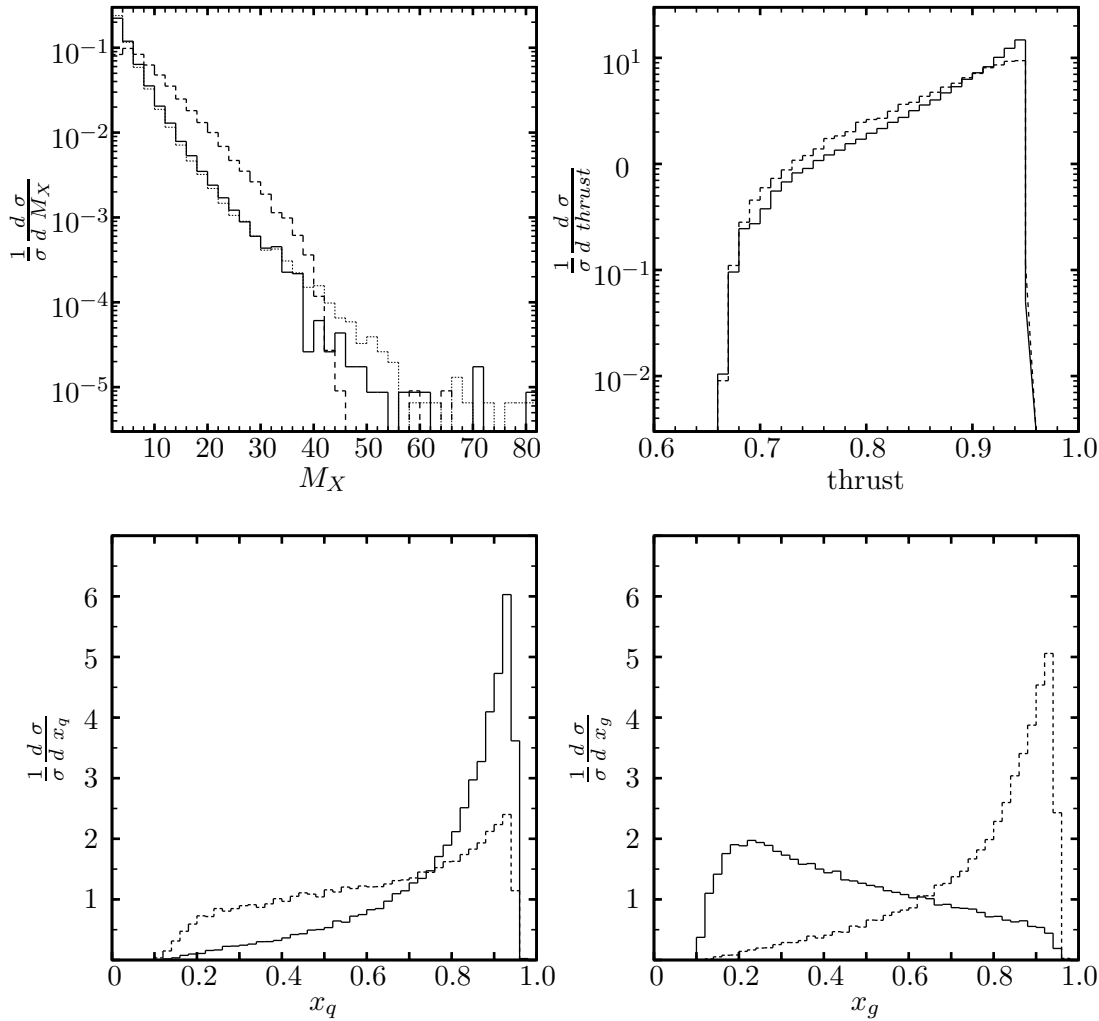
\\ {\sl Figure 3}: The distributions at the parton level. \\ [2mm]
\parbox[]{420pt}{ The solid line refers to the production of three partons, 
using the $Pqq$ vertex. The dashed line corresponds to the $Pgg$ vertex. 
The dotted line in the figure at the top left refers to the production 
of two partons.}
\end{center}
 To deepen our insight into the roles played by gluons
in the two different three-parton processes we are considering, it is
 instructive to look at the parton fractional momenta, defined as usual by  
 $x_q = 2E_q/\Sigma E_{hadr}$ and $x_g = 2E_g/\Sigma 
E_{hadr}$.
They are presented in  
Figures 3c and 3d respectively. It is striking to see that
while bremstrahlung gluons are soft partons 
as expected, those coming from Pomeron splitting are
much harder. 

In order to study to what extent the properties exhibited above are masked
by fragmentation effects, we have calculated the thrust distributions in the
 hadronic centre of mass, after fragmentation, for various ranges in $M_X$.

\begin{center}\begin{picture}(440,400)(0,0)
%
%
\SetOffset(40,225)
\SetWidth{1.0}\BBox(0,0)(160,160)\SetWidth{0.5}
\LinAxis(0,0)(160,0)(5,4,2.5,0,1)
\LinAxis(0,160)(160,160)(5,4,-2.5,0,1)
\LogAxis(0,0)(0,160)(3,-2.5,0,1)
\LogAxis(160,0)(160,160)(3,2.5,0,1)
\Text(1,-6)[]{0.5}
\Text(33,-6)[]{0.6}
\Text(65,-6)[]{0.7}
\Text(97,-6)[]{0.8}
\Text(129,-6)[]{0.9}
\Text(161,-6)[]{1.0}
\Text(80,-18)[]{thrust, $5$GeV$ < M_X < 6.5$GeV}
\Text(-12,160)[]{$10^1$}
\Text(-12,107)[]{$1$}
\Text(-12,53)[]{$10^{-1}$}
\Text(-12,3)[]{$10^{-2}$}
\rText(-28,100)[][l]{$\frac{1}{\sigma}\frac{d\ \sigma}{d\ thrust}$}
%
%
\Line(0.000000,0.000000)(8.000000,0.000000)
\Line(8.000000,0.000000)(16.000000,0.000000)
\Line(16.000000,0.000000)(16.000000,22.630903)
\Line(16.000000,22.630903)(24.000000,22.630903)
\Line(24.000000,22.630903)(24.000000,52.725379)
\Line(24.000000,52.725379)(32.000000,52.725379)
\Line(32.000000,52.725379)(32.000000,92.778659)
\Line(32.000000,92.778659)(40.000000,92.778659)
\Line(40.000000,92.778659)(40.000000,108.833602)
\Line(40.000000,108.833602)(48.000000,108.833602)
\Line(48.000000,108.833602)(48.000000,120.136511)
\Line(48.000000,120.136511)(56.000000,120.136511)
\Line(56.000000,120.136511)(56.000000,130.057060)
\Line(56.000000,130.057060)(64.000000,130.057060)
\Line(64.000000,130.057060)(64.000000,135.433861)
\Line(64.000000,135.433861)(72.000000,135.433861)
\Line(72.000000,135.433861)(72.000000,138.325967)
\Line(72.000000,138.325967)(80.000000,138.325967)
\Line(80.000000,138.325967)(80.000000,141.451523)
\Line(80.000000,141.451523)(88.000000,141.451523)
\Line(88.000000,141.451523)(88.000000,141.221970)
\Line(88.000000,141.221970)(96.000000,141.221970)
\Line(96.000000,141.221970)(96.000000,141.724028)
\Line(96.000000,141.724028)(104.000000,141.724028)
\Line(104.000000,141.724028)(104.000000,138.740516)
\Line(104.000000,138.740516)(112.000000,138.740516)
\Line(112.000000,138.740516)(112.000000,134.342240)
\Line(112.000000,134.342240)(120.000000,134.342240)
\Line(120.000000,134.342240)(120.000000,127.282180)
\Line(120.000000,127.282180)(128.000000,127.282180)
\Line(128.000000,127.282180)(128.000000,120.364717)
\Line(128.000000,120.364717)(136.000000,120.364717)
\Line(136.000000,120.364717)(136.000000,102.417922)
\Line(136.000000,102.417922)(144.000000,102.417922)
\Line(144.000000,102.417922)(144.000000,83.757733)
\Line(144.000000,83.757733)(152.000000,83.757733)
\Line(152.000000,83.757733)(152.000000,50.517766)
\Line(152.000000,50.517766)(160.000000,50.517766)
%
%
\DashLine(0.000000,0.000000)(8.000000,0.000000){2}
\DashLine(8.000000,0.000000)(16.000000,0.000000){2}
\DashLine(16.000000,0.000000)(16.000000,22.139038){2}
\DashLine(16.000000,22.139038)(24.000000,22.139038){2}
\DashLine(24.000000,22.139038)(24.000000,79.695371){2}
\DashLine(24.000000,79.695371)(32.000000,79.695371){2}
\DashLine(32.000000,79.695371)(32.000000,103.817913){2}
\DashLine(32.000000,103.817913)(40.000000,103.817913){2}
\DashLine(40.000000,103.817913)(40.000000,123.345815){2}
\DashLine(40.000000,123.345815)(48.000000,123.345815){2}
\DashLine(48.000000,123.345815)(48.000000,132.243463){2}
\DashLine(48.000000,132.243463)(56.000000,132.243463){2}
\DashLine(56.000000,132.243463)(56.000000,136.764058){2}
\DashLine(56.000000,136.764058)(64.000000,136.764058){2}
\DashLine(64.000000,136.764058)(64.000000,141.367220){2}
\DashLine(64.000000,141.367220)(72.000000,141.367220){2}
\DashLine(72.000000,141.367220)(72.000000,145.205493){2}
\DashLine(72.000000,145.205493)(80.000000,145.205493){2}
\DashLine(80.000000,145.205493)(80.000000,141.232164){2}
\DashLine(80.000000,141.232164)(88.000000,141.232164){2}
\DashLine(88.000000,141.232164)(88.000000,140.683917){2}
\DashLine(88.000000,140.683917)(96.000000,140.683917){2}
\DashLine(96.000000,140.683917)(96.000000,135.927779){2}
\DashLine(96.000000,135.927779)(104.000000,135.927779){2}
\DashLine(104.000000,135.927779)(104.000000,133.788194){2}
\DashLine(104.000000,133.788194)(112.000000,133.788194){2}
\DashLine(112.000000,133.788194)(112.000000,122.449096){2}
\DashLine(112.000000,122.449096)(120.000000,122.449096){2}
\DashLine(120.000000,122.449096)(120.000000,113.209448){2}
\DashLine(120.000000,113.209448)(128.000000,113.209448){2}
\DashLine(128.000000,113.209448)(128.000000,102.413704){2}
\DashLine(128.000000,102.413704)(136.000000,102.413704){2}
\DashLine(136.000000,102.413704)(136.000000,87.762980){2}
\DashLine(136.000000,87.762980)(144.000000,87.762980){2}
\DashLine(144.000000,87.762980)(144.000000,63.640438){2}
\DashLine(144.000000,63.640438)(152.000000,63.640438){2}
\DashLine(152.000000,63.640438)(152.000000,0.000000){2}
\DashLine(152.000000,0.000000)(160.000000,0.000000){2}
%
%
\SetOffset(245,225)
\SetWidth{1.0}\BBox(0,0)(160,160)\SetWidth{0.5}
\LinAxis(0,0)(160,0)(5,4,2.5,0,1)
\LinAxis(0,160)(160,160)(5,4,-2.5,0,1)
\LogAxis(0,0)(0,160)(3,-2.5,0,1)
\LogAxis(160,0)(160,160)(3,2.5,0,1)
\Text(1,-6)[]{0.5}
\Text(33,-6)[]{0.6}
\Text(65,-6)[]{0.7}
\Text(97,-6)[]{0.8}
\Text(129,-6)[]{0.9}
\Text(161,-6)[]{1.0}
\Text(80,-18)[]{thrust, $6.5$GeV$ < M_X < 9$GeV}
\Text(-12,160)[]{$10^1$}
\Text(-12,107)[]{$1$}
\Text(-12,53)[]{$10^{-1}$}
\Text(-12,3)[]{$10^{-2}$}
\rText(-28,100)[][l]{$\frac{1}{\sigma}\frac{d\ \sigma}{d\ thrust}$}
%
%
\Line(0.000000,0.000000)(8.000000,0.000000)
\Line(8.000000,0.000000)(16.000000,0.000000)
\Line(16.000000,0.000000)(16.000000,12.908495)
\Line(16.000000,12.908495)(24.000000,12.908495)
\Line(24.000000,12.908495)(24.000000,66.241828)
\Line(24.000000,66.241828)(32.000000,66.241828)
\Line(32.000000,66.241828)(32.000000,77.840973)
\Line(32.000000,77.840973)(40.000000,77.840973)
\Line(40.000000,77.840973)(40.000000,97.163620)
\Line(40.000000,97.163620)(48.000000,97.163620)
\Line(48.000000,97.163620)(48.000000,107.935453)
\Line(48.000000,107.935453)(56.000000,107.935453)
\Line(56.000000,107.935453)(56.000000,123.012927)
\Line(56.000000,123.012927)(64.000000,123.012927)
\Line(64.000000,123.012927)(64.000000,127.451234)
\Line(64.000000,127.451234)(72.000000,127.451234)
\Line(72.000000,127.451234)(72.000000,133.572555)
\Line(72.000000,133.572555)(80.000000,133.572555)
\Line(80.000000,133.572555)(80.000000,137.785002)
\Line(80.000000,137.785002)(88.000000,137.785002)
\Line(88.000000,137.785002)(88.000000,143.215854)
\Line(88.000000,143.215854)(96.000000,143.215854)
\Line(96.000000,143.215854)(96.000000,144.116275)
\Line(96.000000,144.116275)(104.000000,144.116275)
\Line(104.000000,144.116275)(104.000000,144.276285)
\Line(104.000000,144.276285)(112.000000,144.276285)
\Line(112.000000,144.276285)(112.000000,140.140826)
\Line(112.000000,140.140826)(120.000000,140.140826)
\Line(120.000000,140.140826)(120.000000,135.043675)
\Line(120.000000,135.043675)(128.000000,135.043675)
\Line(128.000000,135.043675)(128.000000,123.990386)
\Line(128.000000,123.990386)(136.000000,123.990386)
\Line(136.000000,123.990386)(136.000000,107.743228)
\Line(136.000000,107.743228)(144.000000,107.743228)
\Line(144.000000,107.743228)(144.000000,90.903056)
\Line(144.000000,90.903056)(152.000000,90.903056)
\Line(152.000000,90.903056)(152.000000,22.300029)
\Line(152.000000,22.300029)(160.000000,22.300029)
%
%
\DashLine(0.000000,0.000000)(8.000000,0.000000){2}
\DashLine(8.000000,0.000000)(16.000000,0.000000){2}
\DashLine(16.000000,0.000000)(16.000000,33.493356){2}
\DashLine(16.000000,33.493356)(24.000000,33.493356){2}
\DashLine(24.000000,33.493356)(24.000000,79.642765){2}
\DashLine(24.000000,79.642765)(32.000000,79.642765){2}
\DashLine(32.000000,79.642765)(32.000000,102.881624){2}
\DashLine(32.000000,102.881624)(40.000000,102.881624){2}
\DashLine(40.000000,102.881624)(40.000000,118.936557){2}
\DashLine(40.000000,118.936557)(48.000000,118.936557){2}
\DashLine(48.000000,118.936557)(48.000000,130.884013){2}
\DashLine(48.000000,130.884013)(56.000000,130.884013){2}
\DashLine(56.000000,130.884013)(56.000000,139.849114){2}
\DashLine(56.000000,139.849114)(64.000000,139.849114){2}
\DashLine(64.000000,139.849114)(64.000000,140.313927){2}
\DashLine(64.000000,140.313927)(72.000000,140.313927){2}
\DashLine(72.000000,140.313927)(72.000000,140.919515){2}
\DashLine(72.000000,140.919515)(80.000000,140.919515){2}
\DashLine(80.000000,140.919515)(80.000000,142.085162){2}
\DashLine(80.000000,142.085162)(88.000000,142.085162){2}
\DashLine(88.000000,142.085162)(88.000000,142.156103){2}
\DashLine(88.000000,142.156103)(96.000000,142.156103){2}
\DashLine(96.000000,142.156103)(96.000000,137.022959){2}
\DashLine(96.000000,137.022959)(104.000000,137.022959){2}
\DashLine(104.000000,137.022959)(104.000000,131.227165){2}
\DashLine(104.000000,131.227165)(112.000000,131.227165){2}
\DashLine(112.000000,131.227165)(112.000000,128.962716){2}
\DashLine(112.000000,128.962716)(120.000000,128.962716){2}
\DashLine(120.000000,128.962716)(120.000000,117.748483){2}
\DashLine(120.000000,117.748483)(128.000000,117.748483){2}
\DashLine(128.000000,117.748483)(128.000000,105.089233){2}
\DashLine(128.000000,105.089233)(136.000000,105.089233){2}
\DashLine(136.000000,105.089233)(136.000000,72.266623){2}
\DashLine(136.000000,72.266623)(144.000000,72.266623){2}
\DashLine(144.000000,72.266623)(144.000000,65.603223){2}
\DashLine(144.000000,65.603223)(152.000000,65.603223){2}
\DashLine(152.000000,65.603223)(152.000000,0.000000){2}
\DashLine(152.000000,0.000000)(160.000000,0.000000){2}
%
%
\SetOffset(40,25)
\SetWidth{1.0}\BBox(0,0)(160,160)\SetWidth{0.5}
\LinAxis(0,0)(160,0)(5,4,2.5,0,1)
\LinAxis(0,160)(160,160)(5,4,-2.5,0,1)
\LogAxis(0,0)(0,160)(3,-2.5,0,1)
\LogAxis(160,0)(160,160)(3,2.5,0,1)
\Text(1,-6)[]{0.5}
\Text(33,-6)[]{0.6}
\Text(65,-6)[]{0.7}
\Text(97,-6)[]{0.8}
\Text(129,-6)[]{0.9}
\Text(161,-6)[]{1.0}
\Text(80,-18)[]{thrust, $9$GeV$ < M_X < 13$GeV}
\Text(-12,160)[]{$10^1$}
\Text(-12,107)[]{$1$}
\Text(-12,53)[]{$10^{-1}$}
\Text(-12,3)[]{$10^{-2}$}
\rText(-28,100)[][l]{$\frac{1}{\sigma}\frac{d\ \sigma}{d\ thrust}$}
%
%
\Line(0.000000,0.000000)(8.000000,0.000000)
\Line(8.000000,0.000000)(16.000000,0.000000)
\Line(16.000000,0.000000)(16.000000,4.434736)
\Line(16.000000,4.434736)(24.000000,4.434736)
\Line(24.000000,4.434736)(24.000000,45.936136)
\Line(24.000000,45.936136)(32.000000,45.936136)
\Line(32.000000,45.936136)(32.000000,55.327671)
\Line(32.000000,55.327671)(40.000000,55.327671)
\Line(40.000000,55.327671)(40.000000,88.072163)
\Line(40.000000,88.072163)(48.000000,88.072163)
\Line(48.000000,88.072163)(48.000000,100.399567)
\Line(48.000000,100.399567)(56.000000,100.399567)
\Line(56.000000,100.399567)(56.000000,107.607982)
\Line(56.000000,107.607982)(64.000000,107.607982)
\Line(64.000000,107.607982)(64.000000,118.728863)
\Line(64.000000,118.728863)(72.000000,118.728863)
\Line(72.000000,118.728863)(72.000000,122.840499)
\Line(72.000000,122.840499)(80.000000,122.840499)
\Line(80.000000,122.840499)(80.000000,132.138830)
\Line(80.000000,132.138830)(88.000000,132.138830)
\Line(88.000000,132.138830)(88.000000,139.582793)
\Line(88.000000,139.582793)(96.000000,139.582793)
\Line(96.000000,139.582793)(96.000000,140.641834)
\Line(96.000000,140.641834)(104.000000,140.641834)
\Line(104.000000,140.641834)(104.000000,146.548978)
\Line(104.000000,146.548978)(112.000000,146.548978)
\Line(112.000000,146.548978)(112.000000,147.911862)
\Line(112.000000,147.911862)(120.000000,147.911862)
\Line(120.000000,147.911862)(120.000000,145.313353)
\Line(120.000000,145.313353)(128.000000,145.313353)
\Line(128.000000,145.313353)(128.000000,134.532481)
\Line(128.000000,134.532481)(136.000000,134.532481)
\Line(136.000000,134.532481)(136.000000,117.706830)
\Line(136.000000,117.706830)(144.000000,117.706830)
\Line(144.000000,117.706830)(144.000000,93.613290)
\Line(144.000000,93.613290)(152.000000,93.613290)
\Line(152.000000,93.613290)(152.000000,29.881203)
\Line(152.000000,29.881203)(160.000000,29.881203)
%
%
\DashLine(0.000000,0.000000)(8.000000,0.000000){2}
\DashLine(8.000000,0.000000)(16.000000,0.000000){2}
\DashLine(16.000000,0.000000)(16.000000,1.286191){2}
\DashLine(16.000000,1.286191)(24.000000,1.286191){2}
\DashLine(24.000000,1.286191)(24.000000,66.910134){2}
\DashLine(24.000000,66.910134)(32.000000,66.910134){2}
\DashLine(32.000000,66.910134)(32.000000,94.522885){2}
\DashLine(32.000000,94.522885)(40.000000,94.522885){2}
\DashLine(40.000000,94.522885)(40.000000,113.305953){2}
\DashLine(40.000000,113.305953)(48.000000,113.305953){2}
\DashLine(48.000000,113.305953)(48.000000,124.579729){2}
\DashLine(48.000000,124.579729)(56.000000,124.579729){2}
\DashLine(56.000000,124.579729)(56.000000,130.958925){2}
\DashLine(56.000000,130.958925)(64.000000,130.958925){2}
\DashLine(64.000000,130.958925)(64.000000,137.750648){2}
\DashLine(64.000000,137.750648)(72.000000,137.750648){2}
\DashLine(72.000000,137.750648)(72.000000,142.164809){2}
\DashLine(72.000000,142.164809)(80.000000,142.164809){2}
\DashLine(80.000000,142.164809)(80.000000,143.550348){2}
\DashLine(80.000000,143.550348)(88.000000,143.550348){2}
\DashLine(88.000000,143.550348)(88.000000,141.575676){2}
\DashLine(88.000000,141.575676)(96.000000,141.575676){2}
\DashLine(96.000000,141.575676)(96.000000,140.803526){2}
\DashLine(96.000000,140.803526)(104.000000,140.803526){2}
\DashLine(104.000000,140.803526)(104.000000,138.752419){2}
\DashLine(104.000000,138.752419)(112.000000,138.752419){2}
\DashLine(112.000000,138.752419)(112.000000,133.706115){2}
\DashLine(112.000000,133.706115)(120.000000,133.706115){2}
\DashLine(120.000000,133.706115)(120.000000,123.775002){2}
\DashLine(120.000000,123.775002)(128.000000,123.775002){2}
\DashLine(128.000000,123.775002)(128.000000,105.768397){2}
\DashLine(128.000000,105.768397)(136.000000,105.768397){2}
\DashLine(136.000000,105.768397)(136.000000,79.280750){2}
\DashLine(136.000000,79.280750)(144.000000,79.280750){2}
\DashLine(144.000000,79.280750)(144.000000,33.396057){2}
\DashLine(144.000000,33.396057)(152.000000,33.396057){2}
\DashLine(152.000000,33.396057)(152.000000,0.000000){2}
\DashLine(152.000000,0.000000)(160.000000,0.000000){2}
%
%
\SetOffset(245,25)
\SetWidth{1.0}\BBox(0,0)(160,160)\SetWidth{0.5}
\LinAxis(0,0)(160,0)(5,4,2.5,0,1)
\LinAxis(0,160)(160,160)(5,4,-2.5,0,1)
\LogAxis(0,0)(0,160)(3,-2.5,0,1)
\LogAxis(160,0)(160,160)(3,2.5,0,1)
\Text(1,-6)[]{0.5}
\Text(33,-6)[]{0.6}
\Text(65,-6)[]{0.7}
\Text(97,-6)[]{0.8}
\Text(129,-6)[]{0.9}
\Text(161,-6)[]{1.0}
\Text(80,-18)[]{thrust, $M_X > 13$GeV}
\Text(-12,160)[]{$10^1$}
\Text(-12,107)[]{$1$}
\Text(-12,53)[]{$10^{-1}$}
\Text(-12,3)[]{$10^{-2}$}
\rText(-28,100)[][l]{$\frac{1}{\sigma}\frac{d\ \sigma}{d\ thrust}$}
%
%
\Line(0.000000,0.000000)(8.000000,0.000000)
\Line(8.000000,0.000000)(16.000000,0.000000)
\Line(16.000000,0.000000)(24.000000,0.000000)
\Line(24.000000,0.000000)(32.000000,0.000000)
\Line(32.000000,0.000000)(32.000000,53.083692)
\Line(32.000000,53.083692)(40.000000,53.083692)
\Line(40.000000,53.083692)(40.000000,64.915624)
\Line(40.000000,64.915624)(48.000000,64.915624)
\Line(48.000000,64.915624)(48.000000,78.530160)
\Line(48.000000,78.530160)(56.000000,78.530160)
\Line(56.000000,78.530160)(56.000000,99.233101)
\Line(56.000000,99.233101)(64.000000,99.233101)
\Line(64.000000,99.233101)(64.000000,107.176516)
\Line(64.000000,107.176516)(72.000000,107.176516)
\Line(72.000000,107.176516)(72.000000,114.755544)
\Line(72.000000,114.755544)(80.000000,114.755544)
\Line(80.000000,114.755544)(80.000000,122.082667)
\Line(80.000000,122.082667)(88.000000,122.082667)
\Line(88.000000,122.082667)(88.000000,129.279672)
\Line(88.000000,129.279672)(96.000000,129.279672)
\Line(96.000000,129.279672)(96.000000,136.825219)
\Line(96.000000,136.825219)(104.000000,136.825219)
\Line(104.000000,136.825219)(104.000000,144.154102)
\Line(104.000000,144.154102)(112.000000,144.154102)
\Line(112.000000,144.154102)(112.000000,148.553052)
\Line(112.000000,148.553052)(120.000000,148.553052)
\Line(120.000000,148.553052)(120.000000,151.763038)
\Line(120.000000,151.763038)(128.000000,151.763038)
\Line(128.000000,151.763038)(128.000000,148.925645)
\Line(128.000000,148.925645)(136.000000,148.925645)
\Line(136.000000,148.925645)(136.000000,131.474199)
\Line(136.000000,131.474199)(144.000000,131.474199)
\Line(144.000000,131.474199)(144.000000,98.155592)
\Line(144.000000,98.155592)(152.000000,98.155592)
\Line(152.000000,98.155592)(152.000000,27.637226)
\Line(152.000000,27.637226)(160.000000,27.637226)
%
%
\DashLine(0.000000,0.000000)(8.000000,0.000000){2}
\DashLine(8.000000,0.000000)(16.000000,0.000000){2}
\DashLine(16.000000,0.000000)(24.000000,0.000000){2}
\DashLine(24.000000,0.000000)(24.000000,10.623760){2}
\DashLine(24.000000,10.623760)(32.000000,10.623760){2}
\DashLine(32.000000,10.623760)(32.000000,70.034072){2}
\DashLine(32.000000,70.034072)(40.000000,70.034072){2}
\DashLine(40.000000,70.034072)(40.000000,91.959482){2}
\DashLine(40.000000,91.959482)(48.000000,91.959482){2}
\DashLine(48.000000,91.959482)(48.000000,106.771701){2}
\DashLine(48.000000,106.771701)(56.000000,106.771701){2}
\DashLine(56.000000,106.771701)(56.000000,117.635277){2}
\DashLine(56.000000,117.635277)(64.000000,117.635277){2}
\DashLine(64.000000,117.635277)(64.000000,128.392957){2}
\DashLine(64.000000,128.392957)(72.000000,128.392957){2}
\DashLine(72.000000,128.392957)(72.000000,134.966548){2}
\DashLine(72.000000,134.966548)(80.000000,134.966548){2}
\DashLine(80.000000,134.966548)(80.000000,138.881556){2}
\DashLine(80.000000,138.881556)(88.000000,138.881556){2}
\DashLine(88.000000,138.881556)(88.000000,144.122933){2}
\DashLine(88.000000,144.122933)(96.000000,144.122933){2}
\DashLine(96.000000,144.122933)(96.000000,145.292815){2}
\DashLine(96.000000,145.292815)(104.000000,145.292815){2}
\DashLine(104.000000,145.292815)(104.000000,146.074602){2}
\DashLine(104.000000,146.074602)(112.000000,146.074602){2}
\DashLine(112.000000,146.074602)(112.000000,143.570983){2}
\DashLine(112.000000,143.570983)(120.000000,143.570983){2}
\DashLine(120.000000,143.570983)(120.000000,135.020477){2}
\DashLine(120.000000,135.020477)(128.000000,135.020477){2}
\DashLine(128.000000,135.020477)(128.000000,122.272912){2}
\DashLine(128.000000,122.272912)(136.000000,122.272912){2}
\DashLine(136.000000,122.272912)(136.000000,90.163050){2}
\DashLine(136.000000,90.163050)(144.000000,90.163050){2}
\DashLine(144.000000,90.163050)(144.000000,45.461760){2}
\DashLine(144.000000,45.461760)(152.000000,45.461760){2}
\DashLine(152.000000,45.461760)(152.000000,0.0){2}
DashLine(152.000000,0.0)(160.000000,0.0){2}
\end{picture} \\
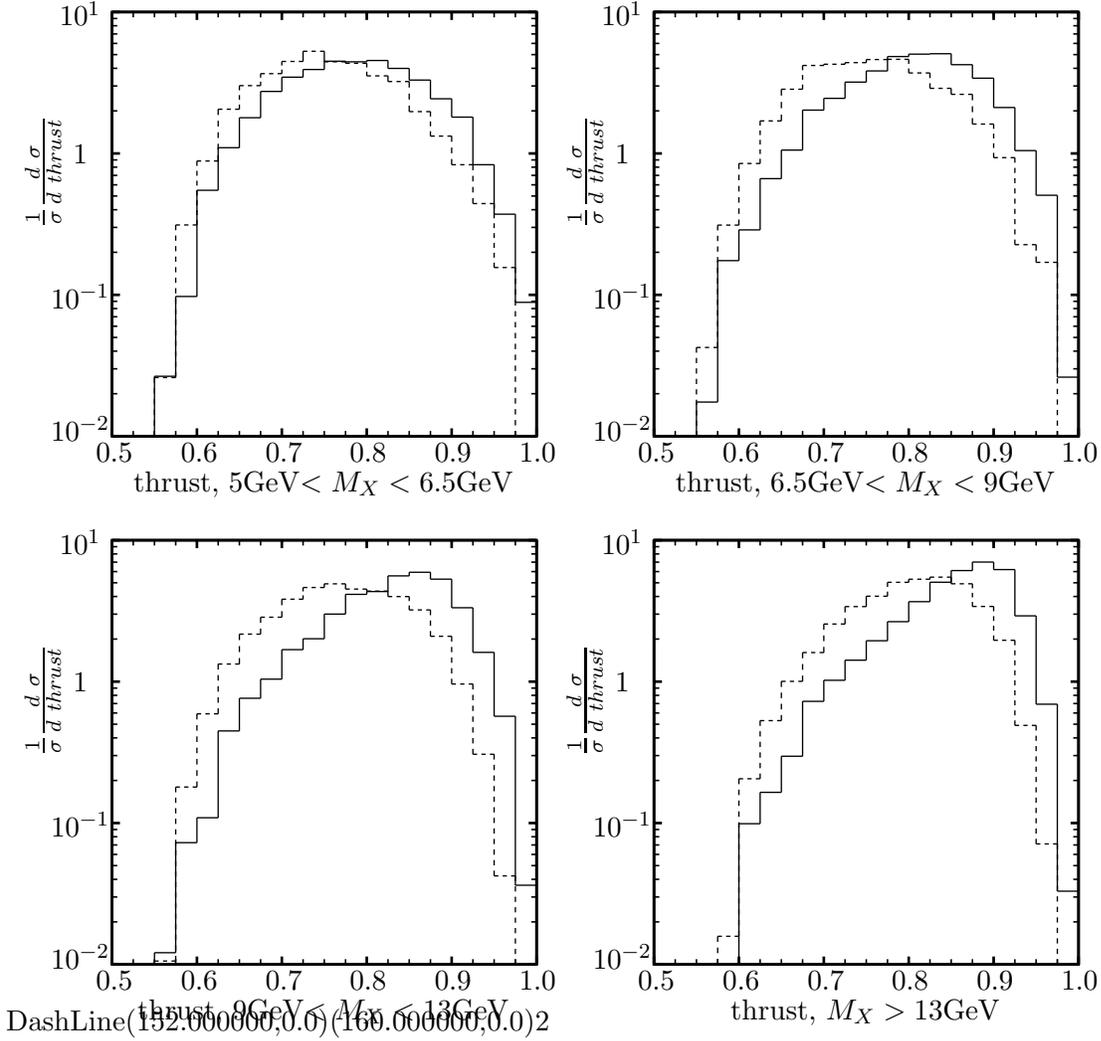
{\sl Figure 4}: The thrust distributions at the hadron level. \\ [2mm]
 \parbox[]{420pt}{ The solid line refers to the production of three partons, 
using the $Pqq$ vertex. The dashed line corresponds to the $Pgg$ vertex.}
\end{center}
The results are shown in Fig. 4. Notice that while at low masses, both thrust
distributions, i.e. the one stemming from the $P \rightarrow q\bar{q}$
coupling and that due to $P\rightarrow g~g$ are similar, at high masses,
the latter gives rise to multihadronic final states which are more spherical 
than the former. \\
To summarize and conclude. We have calculated the cross-sections to 
$O(\alpha_s)$ for diffractive $e-p$ scattering mediated by Pomeron emission.
We have considered a direct coupling of the Pomeron to quark and gluon pairs.
If the Pomeron couples preferentially to gluons as recently indicated by
experimental data \cite{pomg}, the multihadronic final states should exhibit,
as the hadronic
 masses become large, broadening effects much stronger than those
associated to gluon bremstrahlung in $e^+e^-$ annihilation \cite{gluon}.\\

One of the authors (J.V.) likes to thank the Universidad Aut\'onoma in Madrid
for its kind hospitality.

\end{document}